\documentclass[usenatbib]{mn2e}
\newcommand\etal{{et~al.}~}
\newcommand\kms{{~km~s$^{-1}$}}
\newcommand\ih{{~$h_{70}^{-1}$}}
\newcommand\ksmpc{{~$h_{70}$~km~s$^{-1}$~Mpc$^{-1}$}}
\newcommand\vmpc{{~$h_{70}^3$~Mpc$^{-3}$}}

\usepackage{graphicx}
\usepackage{booktabs}
\usepackage{hyperref}

\begin{document}

\title
[Compact groups in theory and practice IV] 
{Compact groups in theory and practice -- IV.  The connection to large-scale structure} 

\author
[Mendel \etal]
{J. Trevor Mendel$^1$\thanks{jtmendel@uvic.ca}, Sara L. Ellison$^1$, Luc Simard$^2$, David R. Patton$^3$ \and Alan W. McConnachie$^2$\\ 
$^1$Department of Physics and Astronomy, University of Victoria, Victoria, British Columbia, V8P 1A1, Canada\\
$^2$National Research Council of Canada, Herzberg Institute of Astrophysics, 5071 West Saanich Road, Victoria, British Columbia, V9E 2E7, Canada\\
$^3$Department of Physics \& Astronomy, Trent University, 1600 West Bank Drive, Peterborough, Ontario, K9J 7B8, Canada}

\maketitle

\begin{abstract}

We investigate the properties of photometrically-selected compact groups (CGs) in the Sloan Digital Sky Survey.  In this paper, the fourth in a series, we focus on understanding the characteristics of our observed CG sample with particular attention paid to quantifying and removing contamination from projected foreground or background galaxies.  Based on a simple comparison of pairwise redshift likelihoods, we find that approximately half of compact groups in the parent sample contain one or more projected (interloping) members; our final clean sample contains 4566 galaxies in 1086 compact groups.  We show that half of the remaining CGs are associated with rich groups (or clusters), i.e. they are embedded sub-structure.  The other half have spatial distributions and number-density profiles consistent with the interpretation that they are either independently distributed structures within the field (i.e. they are isolated) or associated with relatively poor structures.  Comparisons of late-type and red-sequence fractions in radial annuli show that galaxies around apparently isolated compact groups resemble the field population by 300 to 500 kpc from the group centre.  In contrast, the galaxy population surrounding embedded compact groups appears to remain distinct from the field out beyond 1 to 2 Mpc, consistent with results for rich groups.  We take this as additional evidence that the observed distinction between compact groups, i.e. isolated vs. embedded, is a separation between different host environments.

\end{abstract}

\section{Introduction}
\label{sect:intro}

Galaxies' properties are strongly coupled to the characteristics of their surroundings.  The phenomenological distinction between galaxies in high- and low-density environments was established through the discovery of a morphology--density relation \citep{dressler1980,postman1984}, and has since been expanded to incorporate observations showing that galaxies' stellar masses, star-formation rates, mean stellar ages and colours all depend on their surrounding local density \citep[e.g][and references therein]{goto2003,kauffmann2004,balogh2004,blanton2005,cooper2008,bamford2009,skibba2009}.  Despite a wealth of observational evidence, however, the scale(s) at which environment dominates the formation and evolution of galaxies remains unclear.

Disentangling the relative influence of local and large-scale environment relies on understanding the physical processes that can drive galaxy evolution.  In rich groups, both galaxy--galaxy interactions -- e.g. mergers \citep[e.g.][]{toomre1972,mcintosh2008}, tidal interactions and harassment \citep[e.g][]{farouki1981,moore1996} -- and galaxy--environment interactions -- e.g. ram-pressure and viscous stripping \citep{gunn1972,nulsen1982} -- are effective at altering observed properties of the galaxy population, and therefore separating out the particular influence of local or global environment is difficult.  In contrast, the comparatively low encounter velocities and intra-cluster medium densities found in poor groups favour galaxy--galaxy interactions as the dominant pathway for galaxy evolution in these environments.  Study of these systems therefore provides a means of separating out local and global environmental influences.

Compact groups of galaxies (CGs) represent a class of poor groups characterised by both their small number of members and compact angular configurations \citep{rose1977,hickson1982}.  The high densities and low velocity dispersions observed for many CGs \citep{hickson1992,ramella1994} suggest that tidal interactions and mergers should dominate the evolution of galaxies there, making them ideal for the study of these processes.  Initial numerical investigations into the evolution of compact groups suggested that these systems should be short lived, coalescing into a single galaxy over $\sim$Gyr timescales \citep{barnes1985,diaferio1994}.  Reconciling this `fast merger' model with observed CG number densities requires an equally rapid formation of new CGs to replenish those lost in the merging process \citep{diaferio1994}, or the constant feeding of CGs by their surrounding environments \citep{governato1996}.  Alternatively some simulations show that, given particular dynamical configurations, CGs can survive for several Gyrs before merging \citep{governato1991,athanassoula1997}; however, it seems likely that CGs are a natural outcome of the dynamical evolution of initially loose systems \citep[e.g.][]{aceves2002}.

Observational studies of compact group galaxies support the hypothesis that interactions and mergers play a significant role in their evolution, finding that between 40 and 60 per cent show evidence of disturbed photometric or kinematic profiles \citep[e.g.][]{mendes-de-oliveira1994,coziol2007} as well as nuclear (AGN) activity and star formation \citep[e.g.][]{coziol2004,martinez2008}.  \citet{diaferio1994} argue that the constant replenishment of CGs must occur in loose groups, as these environments are favourable to the formation of compact systems on relatively short timescales.  This embedded formation scenario is supported, at least in part, by observational studies showing that $\sim$50 to 70 per cent of compact groups are coincident with other structures \citep[][but see also \citealp{palumbo1995}]{rood1994,barton1998,coziol2004,de-carvalho2005,andernach2007}.  A connection between CGs and larger, rich systems is further supported by a comparison of CG and cluster galaxy populations, which are found to be similar both in terms of morphological mix \citep{hickson1982,zepf1991a,lee2004} and stellar populations \citep{proctor2004a,de-la-rosa2007}.

It seems clear that further understanding of the compact group environment can best be gained from the study of large samples of CGs; however, the relatively low space density of these systems dictates that large survey volumes are required to obtain statistically meaningful samples.  The best known and most well studied catalogue of CGs is that of \citet{hickson1982}, containing 100 groups (HCGs) selected from the Palomar Observatory Sky Survey (POSS).  Subsequent large catalogues of CGs have been published using the Digitized Second Palomar Observatory Sky Survey \citep[DPOSS-II;][]{iovino2003,de-carvalho2005}, CfA Redshift Survey \citep{barton1996} and Sloan Digital Sky Survey \citep[SDSS;][]{lee2004,mcconnachie2009}.  \citet{barton1996} use a friends-of-friends algorithm \citep{huchra1982} to select CGs from within the CfA Redshift Survey, extracting a catalogue of 89 compact systems with 3 or more redshift-confirmed members.  \citet{iovino2003} and \citet{de-carvalho2005} used a modified version of the \cite{hickson1982} selection criteria to identify a total of 459 CGs from $\sim$6260 deg$^2$ of DPOSS-II imaging.  \citet{lee2004} adopt the selection criteria of \citet{iovino2003} to identify CGs from the SDSS Early Data Release catalogues, identifying 175 CGs over 153 deg$^2$ of imaging.  

In this series of papers we focus on a joint observational and theoretical analysis of compact groups, the ultimate aim of which is to highlight the role of CGs in galaxy evolution.  In \citet[hereafter Paper I]{mcconnachie2008}, we used a mock galaxy catalogue from the Millennium Simulation \citep{springel2005,de-lucia2007} to study the spatial properties of compact groups.  This analysis showed that a relatively low fraction, $\sim$30 per cent, of photometrically identified CGs are genuinely compact in three dimensions while the remaining $\sim$70 per cent are partially or entirely comprised of interloping galaxies.  In \citet[hereafter Paper II]{Brasseur2009}, we used our mock catalogue to show that these interloping galaxies are predicted to be both bluer and more heavily star forming than populations of genuine CG galaxies, suggesting that careful accounting of interlopers is key to understanding galaxy evolution in the compact group environment.  Taken together, Papers I and II highlight the heterogeneous nature of compact groups selected using purely photometric criteria.

In \citet[hereafter Paper III]{mcconnachie2009} we adopt the selection criteria outlined by \citet{hickson1982} to identify 2297 CGs from 8417 deg$^2$ of imaging in the SDSS Data Release 6.  This sample represents the largest homogeneously-selected catalogue of CGs currently available, and serves as the basis for the present work.  We describe the sample in greater detail in Section \ref{data}, with particular attention paid to the removal of projected (false) groups.  In Section \ref{physical} we examine the spatial relationship between compact groups and other structures in the SDSS.  This spatial comparison leads us to identify a sub-sample of compact groups that are clearly associated with observed large-scale structure, both in terms of projected separations and surrounding galaxy number-density.  In Sections \ref{galaxy_properties} and \ref{cg_properties} we show that separating compact groups based on their proximity to observed large-scale structure results in two samples with differing photometric and morphological properties.  Throughout this paper we adopt a concordance cosmology with $\Omega_{\Lambda} = 0.7$, $\Omega_{\mathrm{M}} = 0.3$ and $H_0 = 70$\ksmpc.

\section{Data}
\label{data}

In Paper III we describe the photometric selection of compact groups from the SDSS Data Release 6 \citep{adelman-mccarthy2008}, which included imaging of the entire SDSS-II Legacy Survey area.  Since that paper, SDSS Data Release 7 \citep[DR7;][]{abazajian2009} has provided an additional $\sim$1200 deg$^2$ of spectroscopic data, completing spectroscopic observations of the SDSS-II Legacy Survey footprint.  In what follows we use galaxy catalogues drawn from SDSS DR7 and, where available, supplement the CG samples in Paper III with updated spectroscopic information.

\subsection{Galaxy sample}

Much of our analysis relies on the use of volume-limited samples to trace the characteristics of galaxies in and around compact groups.  The parent catalogue for these samples includes all primary photometric objects in DR7 identified as galaxies by the SDSS {\sc photo} pipeline ({\tt photoObj.type = 3}) with extinction-corrected $r$-band Petrosian magnitudes in the range $14.5 < m_r \leq 18.0$.  These limits are chosen to match the selection criteria of the compact group catalogue described in Paper III (see Section \ref{cgs}).  We further remove objects flagged either {\tt DEBLENDED\verb|_|AS\verb|_|PSF} or {\tt SATURATED} to eliminate unresolved and saturated objects from the sample.

Although spectroscopic information for compact group galaxies is limited by fibre collisions at small angular separation, our analysis still benefits from the inclusion of accurate redshift information where available.  We select spectroscopic targets from the photometric sample by requiring that objects have a unique, science-worthy spectrum ({\tt specPhoto.sciencePrimary = 1}) and a redshift confidence greater than 70 per cent ({\tt specPhoto.zConf $\ge 0.7$}).  The resulting sample contains 1\,116\,866 galaxies, 691\,163 of which have spectroscopic information available.

\citet{simard2011} provide an updated catalogue of SDSS photometry for galaxies in DR7, which we adopt throughout this study.  Briefly, \citeauthor{simard2011} fit several different photometric models of varying complexity to the SDSS using {\sc gim2d} \citep{simard2002}, from which we adopt their de Vaucouleurs bulge (S\'ersic $n=4$) plus exponential disk (S\'ersic $n=1$) fits.  These PSF-convolved model decompositions are performed simultaneously in the $g$- and $r$-bands using re-estimated sky background levels and object deblending.  Extinction corrected rest-frame quantities for the best-fit models are computed at $z=0$ using {\sc k-correct v4\verb|_|2} \citep{blanton2007a}.  We adopt these catalogues preferentially over SDSS photometry for the remainder of this paper based on their improved reliability at small angular separations \citep[see][]{simard2011,patton2011}, which is critical to the study of galaxy properties in high-density environments.

\subsection{Compact groups}
\label{cgs}

In Paper III we describe the selection of two CG catalogues, A and B, defined on flux-limited samples with extinction-corrected $r$-band Petrosian magnitudes in the range $14.5 < m_r \leq 18.0$ and $14.5 < m_r \leq 21.0$, respectively.  In this work we adopt Catalogue A, which contains a total of 9713 galaxies in 2297 compact groups.  Briefly, CGs within the flux-limited sample were selected following the criteria outlined by \citet{hickson1982}, such that identified groups satisfy\\

\hangindent=0.5cm
\hangafter=0
\noindent (i) $N \geq 4$,\\
(ii) $\theta_N \geq 3\theta_G$,\\
(iii) $\mu_G < 26.0$ mag arcsec$^{-2}$,\\

\noindent where $N$ is the number of galaxies within 3 mag of the brightest galaxy, $\mu_G$ is the total magnitude of these $N$ galaxies averaged over the minimum circle of angular diameter $\theta_G$ which contains their geometric centres and $\theta_N$ is the maximum (concentric) angular diameter of a circle containing no other galaxies within the group magnitude range or brighter.  Taken together, criterion (i) selects groups of galaxies, while criteria (ii) and (iii) ensure that these groups are sufficiently compact and isolated so as to exclude cluster cores.   As a final step, all groups in Catalogue A were visually inspected, and any galaxy associations identified as a result of gross photometric error or misclassification (e.g. stellar sources, see figure 1 of Paper III) were removed from the final sample.

\subsubsection{Interloper removal}
\label{interlopers}

There is known to be a high incidence of interloping galaxies in photometrically-identified catalogues of poor groups; estimates of the contamination rate -- the fraction of chance projections relative to the total number of catalogued groups -- range from $\sim$30 to 80 per cent depending on classification method and survey magnitude limits \citep[see e.g.][Paper I]{ramella1997,niemi2007,diaz-gimenez2010}.  In Paper III we use the available (DR6) spectroscopic redshifts to demonstrate that the contamination rate for groups in Catalogue A is at least 55 per cent, in agreement with predictions based on semi-analytic models (e.g. figure 7 of Paper III; Paper I).  Given our aim of understanding the properties of compact groups, it is crucial that we attempt to account for these contaminants in our CG catalogue.

Physical limitations on the placement of SDSS fibres prohibit the observation of significant numbers of CG members on a single spectroscopic plate;  the minimum SDSS fibre separation of 55$^{\prime\prime}$ is of order half the median angular diameter $\theta_G$ of groups in Catalogue A.  Consequently the fraction of Catalogue A galaxies with spectra, and hence secure redshifts, is only $\sim$50 per cent (increased from $\sim$43 per cent in Paper III with the inclusion of additional spectroscopy from DR7).  For the remaining 50 per cent of galaxies we adopt photometric redshifts estimated by the SDSS pipeline.  Owing to the significant uncertainty of photometric redshifts relative to the maximum expected line-of-sight separation of CG galaxies \citep[$\sim$1000\kms;][]{ramella1994,hickson1997} it is difficult to explicitly confirm the physical association of a given CG; however, in many cases we can reliably reject systems with discordant spectroscopic or photometric redshifts.  

\begin{table}
\centering
\scriptsize
\begin{tabular}{@{}lcccc@{}} \toprule
	& \multicolumn{4}{c}{Number of photo-zs} \\
	& $N=1$	& $N=2$	& $N=3$	& $N=4$ \\ \cmidrule(l){2-5}
Initial number of groups			& 452	& 777	& 388	& 98   \\ \midrule
& \multicolumn{4}{c}{$\mathcal{L}\ge 0.0027$}\\ \cmidrule(l){2-5}
Number of `clean' groups 		& 163	& 363	& 256	& 64 \\
Estimated false positive rate	& 7.2\%	& 18.7\%	& 43.9\%	& 59.7\% \\
Estimated false negative rate	& 1.3\%	& 2.5\%	& 3.8\%	& 2.5\% \\ \midrule
& \multicolumn{4}{c}{$\mathcal{L}\ge 0.01$}\\ \cmidrule(l){2-5}
Number of `clean' groups 		& 155	& 340	& 220	& 57 \\
Estimated false positive rate	& 6.4\%	& 15.6\%	& 34.2\%	& 42.1\% \\
Estimated false negative rate	& 4.4\%	& 7.9\%	& 10.6\%	& 5.0\% \\ \midrule
& \multicolumn{4}{c}{$\mathcal{L}\ge 0.03$}\\ \cmidrule(l){2-5}
Number of `clean' groups 		& 147	& 289	& 161	& 47 \\
Estimated false positive rate	& 5.3\%	& 12.0\%	& 21.5\%	& 27.2\% \\
Estimated false negative rate	& 9.4\%	& 18.3\%	& 26.9\%	& 25.0\% \\ \midrule
& \multicolumn{4}{c}{$\mathcal{L}\ge 0.05$}\\ \cmidrule(l){2-5}
Number of `clean' groups 		& 135	& 238	& 121	& 36 \\
Estimated false positive rate	& 4.4\%	& 8.5\%	& 13.2\%	& 14.0\% \\
Estimated false negative rate	& 21.9\%	& 40.0\%	& 53.8\%	& 53.0\% \\ \bottomrule
\end{tabular}
\footnotesize
\caption{Contamination estimates for 4 member compact groups.  Columns show the number of 4 member groups remaining in the cleaned sample separated by the number of photometric redshifts they contain for different cuts in pairwise redshift likelihood.}
\label{tab:cont_fraction} 
\end{table}	

Our methodology is as follows.  For each putative CG we compute the pairwise likelihood that two galaxies within the group have a velocity separation of 1000\kms~or less, e.g. for a group of four members there are six such pairwise likelihoods.   We compute the velocity separation likelihood of a given pair assuming Gaussian error distributions for both spectroscopic and photometric redshifts, and consider as unlikely any group whose least-bound pair has an integrated likelihood $\mathcal{L} < 0.0027$ between $\pm$1000\kms, corresponding to a rejection of our velocity separation criteria at greater than $3\sigma$ significance.  For groups with exactly four members the rejection of a single galaxy pairing is sufficient to remove the group from our final CG sample as it fails to meet the CG richness criterion discussed above.  In groups with more than four members, we remove the galaxy that results in the highest minimum pairwise likelihood among remaining group members.   This removal process is iterated until the minimum pairwise likelihood is greater than 0.0027 or the number of remaining CG members falls below four.  

\begin{figure}
\centering
\includegraphics[scale=0.68]{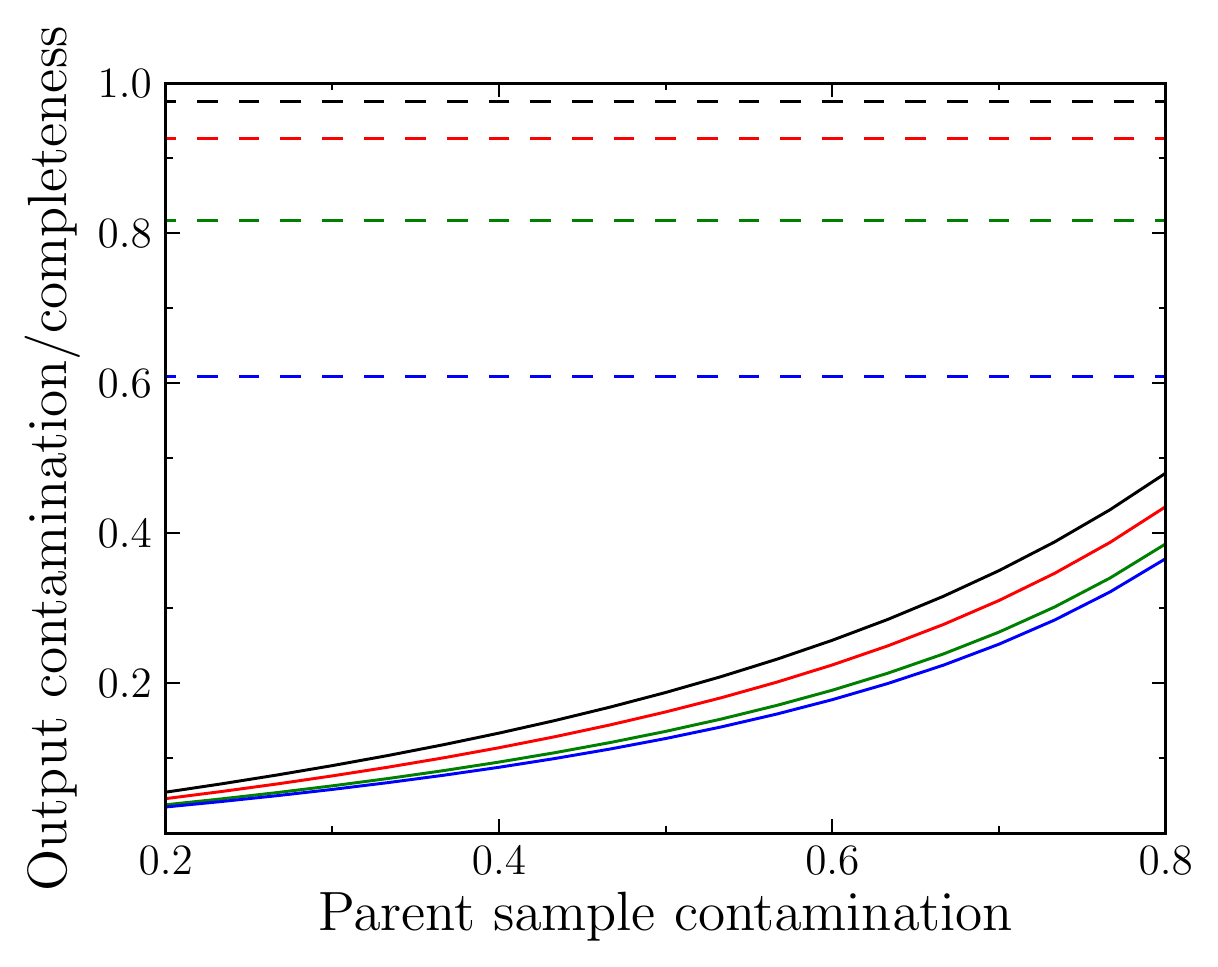}
\caption{Characteristics of the sample cleaning procedure.  Solid lines show the output sample contamination for a given input contamination fraction estimated using mix of observed photometric and spectroscopic redshifts in conjunction with the false positive and negative rates show in Table \ref{tab:cont_fraction}.  Dashed lines show the estimated fraction of genuine groups remaining after interloper removal -- i.e. completeness -- accounting for false negatives in the interloper removal process.  Black, red, green and blue lines (top to bottom) correspond to different pairwise redshift likelihood cuts of $\mathcal{L}=$ 0.0027, 0.01, 0.03 and 0.05 as described in Section \ref{interlopers}.}
\label{fig:contamination}
\end{figure}

We gauge the reliability of this removal process using the subset of 154 four-member CGs where all members have spectroscopic redshifts available -- hereafter referred to as spectroscopic groups.  Four-member groups make up the majority (83 per cent) of the Catalogue A sample, and we choose to focus on these CGs for our reliability estimates as the presence of a single interloping galaxy removes these systems from our final sample, allowing for a straight-forward interpretation of the contamination rate.  For each spectroscopic group we first assess the likelihood of it being a genuine association following the probabilistic procedure described above.  Given the small redshift error relative to our expected velocity spread of 1000\kms  (the median spectroscopic redshift error for the CG sample corresponds to $c\Delta z \approx 50$\kms) this measurement is unambiguous for the majority of groups.  Using this initial likelihood to separate genuine and projected groups, we then randomly replace some number of group members' spectroscopic redshifts with their corresponding photometric redshifts and recompute the distribution of pairwise likelihoods.  In this way we can estimate the probability that a genuine group will be rejected by our interloper removal scheme (false negative), or alternatively that a projected `group' will be wrongly accepted (false positive), given some arbitrary mix of spectroscopic and photometric redshifts.  We also recompute the CG catalogue using several different redshift likelihood cuts to assess the robustness of our adopted limits.  The results of this comparison are summarised in Table \ref{tab:cont_fraction}.  

\begin{figure}
\centering
\includegraphics[scale=0.68]{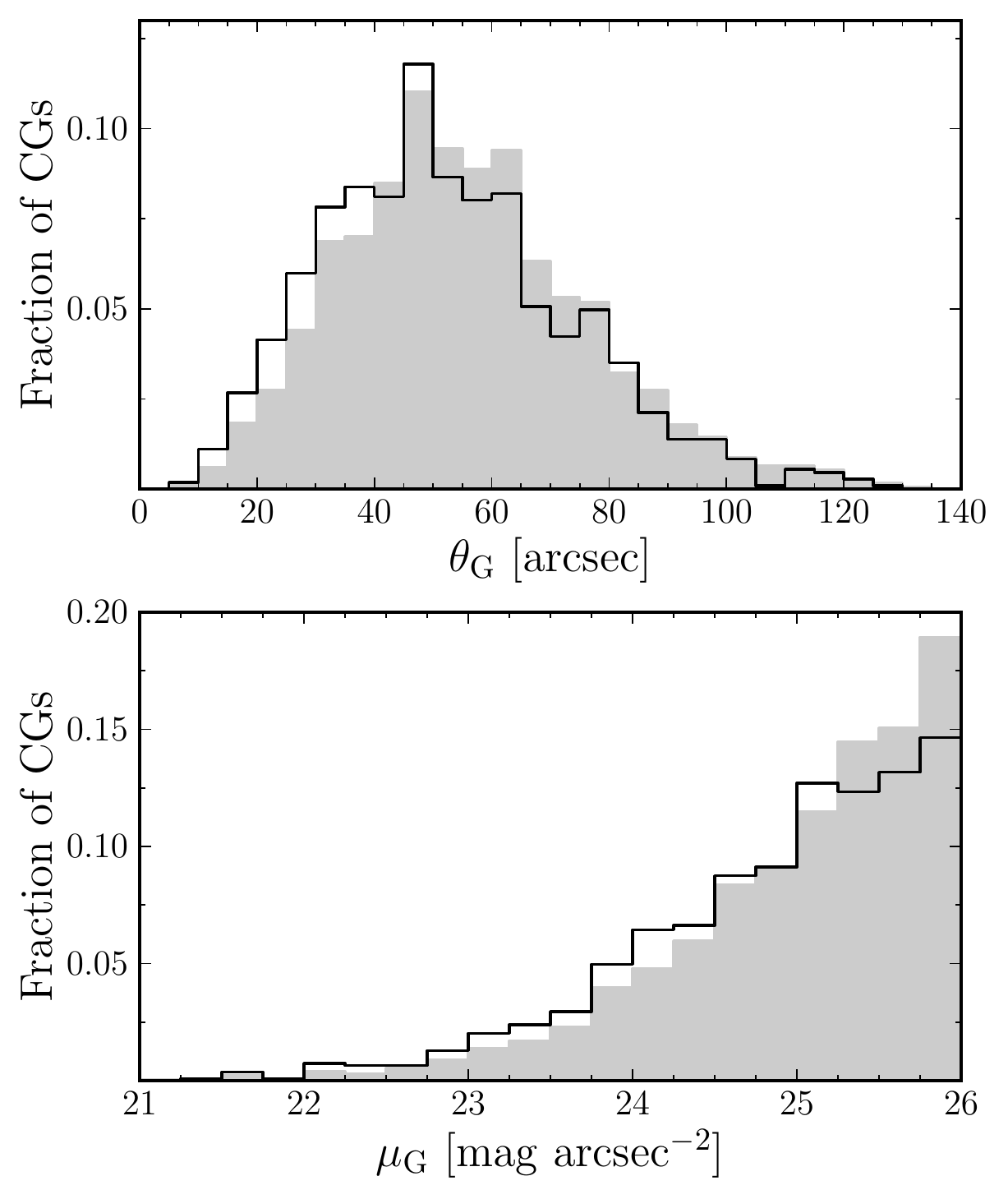}
\caption{{\it Top panel}: Distribution of compact group angular sizes, $\theta_\mathrm{G}$, for the initial Catalogue A (shaded histogram) and the final sample after removal of likely interlopers (open histogram, see Section \ref{interlopers}).  Groups with the largest apparent angular sizes are preferentially removed as being chance projections.  {\it Bottom panel}:  Distribution of group surface brightnesses, $\mu_\mathrm{G}$, before and after interloper removal.  Shading is the same as above.}
\label{fig:mu_theta}
\end{figure}

\begin{figure}
\centering
\includegraphics[scale=0.68]{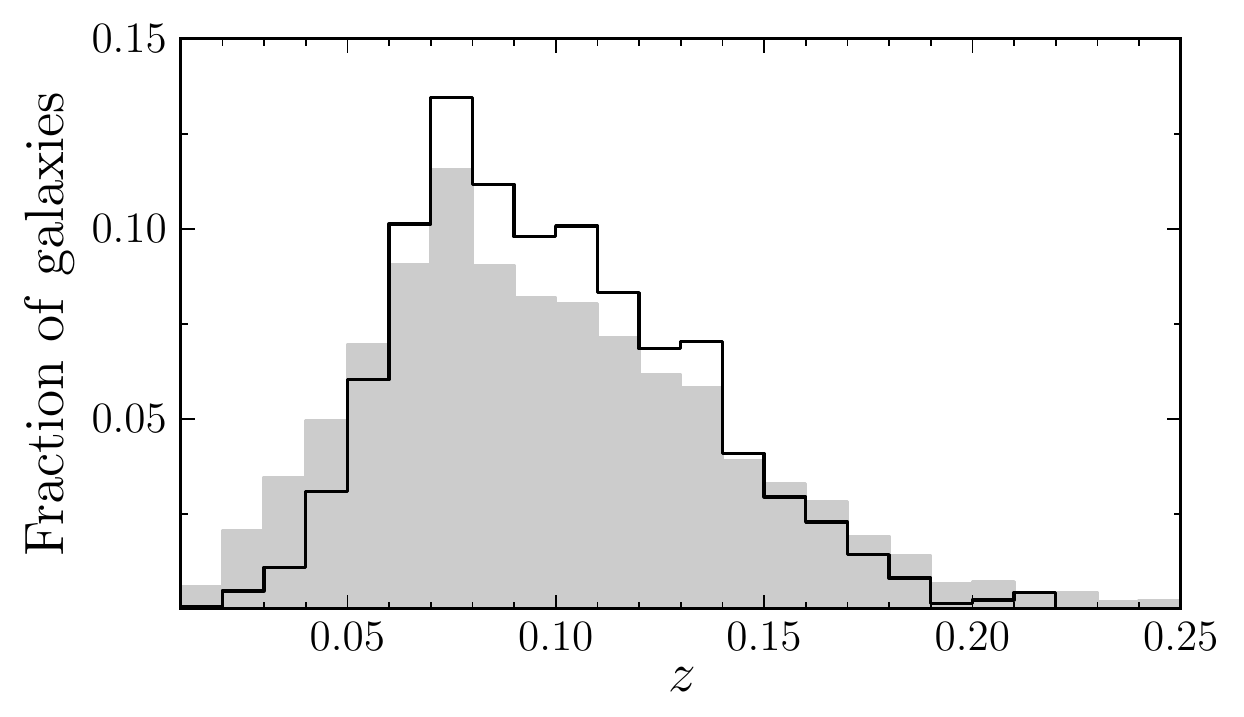}
\caption{Distribution of compact group galaxy redshifts for the input catalogue from Paper III (shaded histogram) and the cleaned CG sample (open histogram).  Chance superpositions are found to preferentially occupy the low probability tails of the redshift distribution at high and low $z$.}
\label{fig:cg_redshifts}
\end{figure}

In general, our relatively tolerant likelihood requirements mean that we reject a minimum of genuine groups while still removing at least 40 per cent of spurious associations, even when no galaxies in the group have spectroscopic redshifts available.  We can translate the results in Table \ref{tab:cont_fraction} into an estimate of the remaining contamination in the `cleaned' CG sample assuming a value for the inherent Catalogue A contamination rate.  We show in Figure \ref{fig:contamination} the predicted properties of the cleaned sample for a range of input contamination rates from 20 to 80 per cent.  Assuming an input catalogue contamination of 70 to 75 per cent (Paper III) and our adopted likelihood limit of 0.0027, we estimate that as much as 35 to 40 per cent contamination may remain, comparable to that expected from using a surface-brightness cut of $\mu_G < 24$~mag~arcsec$^{-2}$ (Paper III).  Although more stringent pairwise likelihood cuts can be used to further decrease the sample contamination (down to $\sim$20 per cent using a cut of $\mathcal{L} \ge 0.05$), more stringent cuts exact an increasingly heavy toll on the remaining fraction of genuine groups.  We therefore adopt a sample selected using a relatively loose cut of $\mathcal{L} \ge 0.0027$ for the remainder of the paper to maximise the final group sample, but we will revisit the potential influence of the remaining interlopers in Section \ref{cg_properties}.

All told, the procedure described above removes 1211 groups from Catalogue A, resulting in a final sample of 1086 CGs.  In Figure \ref{fig:mu_theta} we show a comparison of group angular size and surface brightness before and after removal of interloping groups.  As expected, we find that projected systems have preferentially larger sizes and lower surface brightnesses than associations that are more likely to be genuine CGs \citep[e.g.][Paper I]{diaz-gimenez2010}.  In Figure \ref{fig:cg_redshifts} we show the comparison of CG galaxy redshifts before and after the interloper removal process.  Rejection of interloping galaxies preferentially removes galaxies from the low-probability tails of the redshift distribution at both high- and low-$z$.

Although we identify CGs over the entire SDSS Legacy Survey footprint, we exclude here the three isolated stripes in the Southern Galactic Cap (stripes 76, 82 and 86) {and a small isolated area in the Northern Galactic Cap (stripes 42, 43 and 44)} to match the selection of our rich group sample (see below) and limit potential edge effects in our radial analyses.  We additionally impose a redshift cut of $0.01 \le z_\mathrm{cg} \le 0.14$ to match the limits of our rich group catalogue, described below.  The redshift and spatial cuts described above, as well as our removal of projected systems, results in a final sample of 3455 galaxies in 819 compact groups with a median redshift $z = 0.09$.  A table of cleaned CGs and listing of their individual galaxies is provided in Appendix \ref{tables} and available online.

\subsection{Rich groups from \citealp{tago2010}}
\label{clusters}

Much of our analysis relies upon the comparison of CG environments with those of other, known structures in the SDSS.  We adopt for this purpose the flux-limited group catalogue published by \citet[hereafter T10]{tago2010} , which uses as its input the SDSS spectroscopic main galaxy sample.  T10 identify groups in the contiguous region of the SDSS Northern Galactic Cap through application of a friends-of-friends algorithm \citep[FoF;][]{huchra1982} which, starting from individual galaxies, `grows' groups through the inclusion of additional galaxies satisfying pre-determined projected and velocity separation criteria (linking lengths).  T10 use a redshift-dependent linking length tuned to recover artificially-redshifted rich groups, and show that with this approach they reliably identify these mock groups up to $z \approx 0.14$, above which their catalogue is likely to become increasingly inhomogeneous.  We therefore adopt a redshift range of $0.01 \le z \le 0.14$ throughout for both our group and CG samples, where the lower redshift limit is imposed to avoid galaxies in the Local Supercluster.
 
The FoF algorithm identifies any and all `bound' groups (and clusters) down to a limiting richness of $N = 2$.  Our interest in adopting the T10 group catalogue as a comparison sample comes from the desire to compare the properties of CGs with more massive structures; as a number of our groups contain 2-3 spectroscopic redshifts, it makes sense to exclude these poor systems from the T10 catalogue.  We therefore limit the adopted T10 groups to a richness $N > 4$.  The remaining sample contains 15\,361 groups, which we will hereafter refer to as rich groups for brevity, but which includes structures up to the cluster scale.  

As is the case with any flux-limited galaxy sample, it becomes difficult to reliably select groups as redshift increases and the range of observed {\it absolute} magnitude falls.  In particular, the flux-limited nature of the input SDSS catalogue dictates that the spatial number-density of groups will decrease with increasing redshift independent of astrophysical effects relating to the growth of structure (cf. figures 2 and 6 of \citealp{tago2010}), and we must therefore exercise care in interpreting the comparison of CGs and rich groups.

\subsubsection{Assessing incompleteness in the T10 catalogue}
\label{group_incomp}

We are interested in comparing the relative spatial distributions of CGs and rich groups; however before doing so it is useful to consider how structural incompleteness can enter in to the T10 catalogue.  In instances where only a single spectroscopic redshift exists for a group member -- due to, for example, limitations in the SDSS fibre placement or the flux limit of the parent sample -- that galaxy will be identified as isolated by the FoF algorithm.  Alternatively, if the typical galaxy--galaxy separation in a given group exceeds the adopted FoF linking length criteria, group members will be incorrectly identified as isolated.

The latter point, that galaxy separations exceed the adopted linking lengths, is beyond the scope of this paper to assess.  The former, however, can be understood by considering the properties of galaxy samples which include both spectroscopic and photometric targets (as opposed to the purely spectroscopic sample used by T10); we adopt a relatively simple approach to estimate the contribution from this type of incompleteness.  For each group in the T10 catalogue, we compute what will hereafter referred to as the minimum photometric richness, $n_\mathrm{min}$, which we define as the number of galaxies within a linking volume (that is, the volume defined by the radial and transverse linking length criteria at a given redshift) about the group centre.  We use as the catalogue for this measurement the parent photometric catalogue described in Section \ref{data} -- i.e. the 1\,116\,866 galaxies satisfying our adopted magnitude limits and quality criteria.  For comparison, we also measure the minimum photometric richness about isolated galaxies in the T10 sample (i.e. those galaxies that are not found to have nearby spectroscopic neighbours by T10).  For any group or galaxy, $n_{min}$ therefore carries information about the scale of the group that {\it could} have been found given perfect spectroscopic completeness.

\begin{figure}
\centering
\includegraphics[scale=0.68]{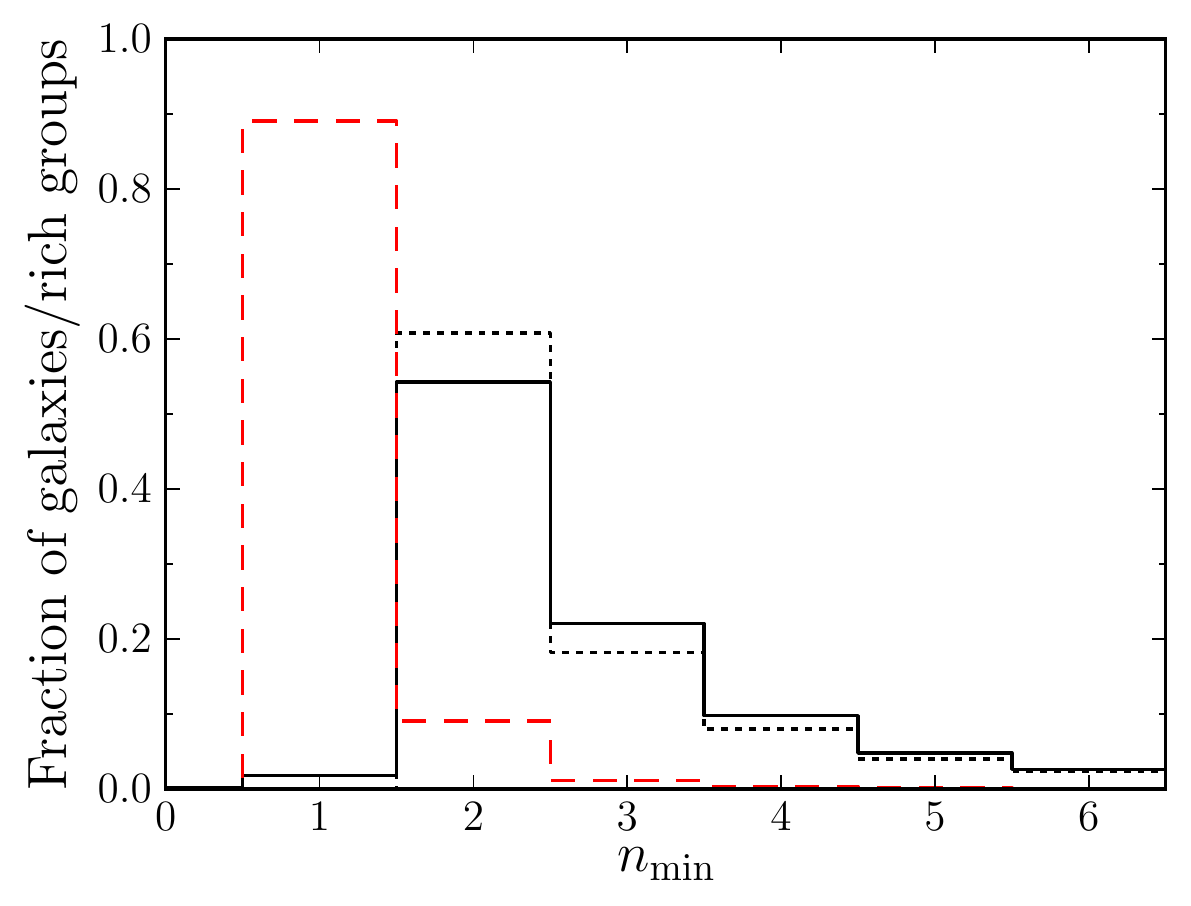}
\caption{Distributions of minimum photometric richness, $n_\mathrm{min}$ (see Section \ref{group_incomp}), for `isolated' galaxies and groups in the T10 catalogue. Solid (black) and dashed (red) histograms show the distributions for groups and isolated galaxies, respectively.  The dotted (black) histogram shows the distribution of total group (spectroscopic) richness quoted by \citet{tago2010}.}
\label{fig:nmin}
\end{figure}

T10 adopt a minimum transverse linking length, $LL_{p0}$, of $\sim$357\ih~kpc at their minimum redshift ($z = 0.009$), which scales as  $1+\arctan(\Delta z/0.05)$.  The ratio between transverse and radial linking lengths  ($LL_{r0}/LL_{p0}$) is 10 (consistent with previous implementations of FoF group finders, e.g., \citealp{eke2004}), such that with the minimum projected linking criterion of 357\ih~kpc, $LL_{r0} = 250$\kms.  In Figure \ref{fig:nmin} we plot the distribution of $n_{min}$ for rich groups (black histogram) and isolated galaxies (red histogram).  Note that we do not account for multiplicity of $n_\mathrm{min}$ in the isolated galaxy sample, i.e. $n_\mathrm{min}$ represents the minimum photometric richness for each galaxy, as opposed to each structure as it does for the rich groups.  The distribution of $n_\mathrm{min}$ for `isolated' galaxies suggests that the majority ($\sim$90 per cent) are genuinely isolated given the adopted flux limit (i.e. $m_r \leq 18$), and we therefore do not expect this type of incompleteness to significantly effect our results.  If we consider structures above our adopted richness limit of 4 (Section \ref{clusters}), spectroscopic incompleteness contributes at less than the 1 per cent level.

\section{The global environments of compact groups}
\label{physical}

We are interested in understanding not only the galaxy content of compact groups but the environments in which they are found.  In this section, we aim to address both the spatial distribution of CGs relative to other large-scale structures, as well as the physical reality of groups in the cleaned Catalogue A.

\subsection{Physical proximity to large-scale structure}
\label{corr}

Structure formation in a $\Lambda$CDM cosmology occurs preferentially in regions of high density, and it may therefore be expected that CGs are closely related to other collapsed structures.  Previous studies of CGs have assessed their spatial coincidence following visual examination or cross-correlation with other catalogues of large-scale structures.  The environments of CGs in the \citet{hickson1982} catalogue have been assessed by \citet{ramella1994} and \citet{rood1994}, both of whom find that $\sim$70 per cent of HCGs are coincident with loose groups and clusters; similar conclusions are reached by \citet{andernach2005}.  \citet{de-carvalho2005} use their significantly larger sample of 459 CGs to show that $\sim$66 per cent of their groups have at least one other catalogued structure nearby, the majority of which are clusters or loose groups (97 per cent).  In Paper I we adopt a theoretical approach to the analysis of CG environments, classifying compact associations (CAs) selected from the semi-analytic models of \citet{de-lucia2007} based on their occupation of dark matter haloes.  In that work, we show that approximately half of CAs are found occupying a single dark matter halo containing few (if any) other galaxies; remaining CAs are either found occupying single haloes with multiple other galaxies or are spread across several haloes.

\begin{figure}
\centering
\includegraphics[scale=0.68]{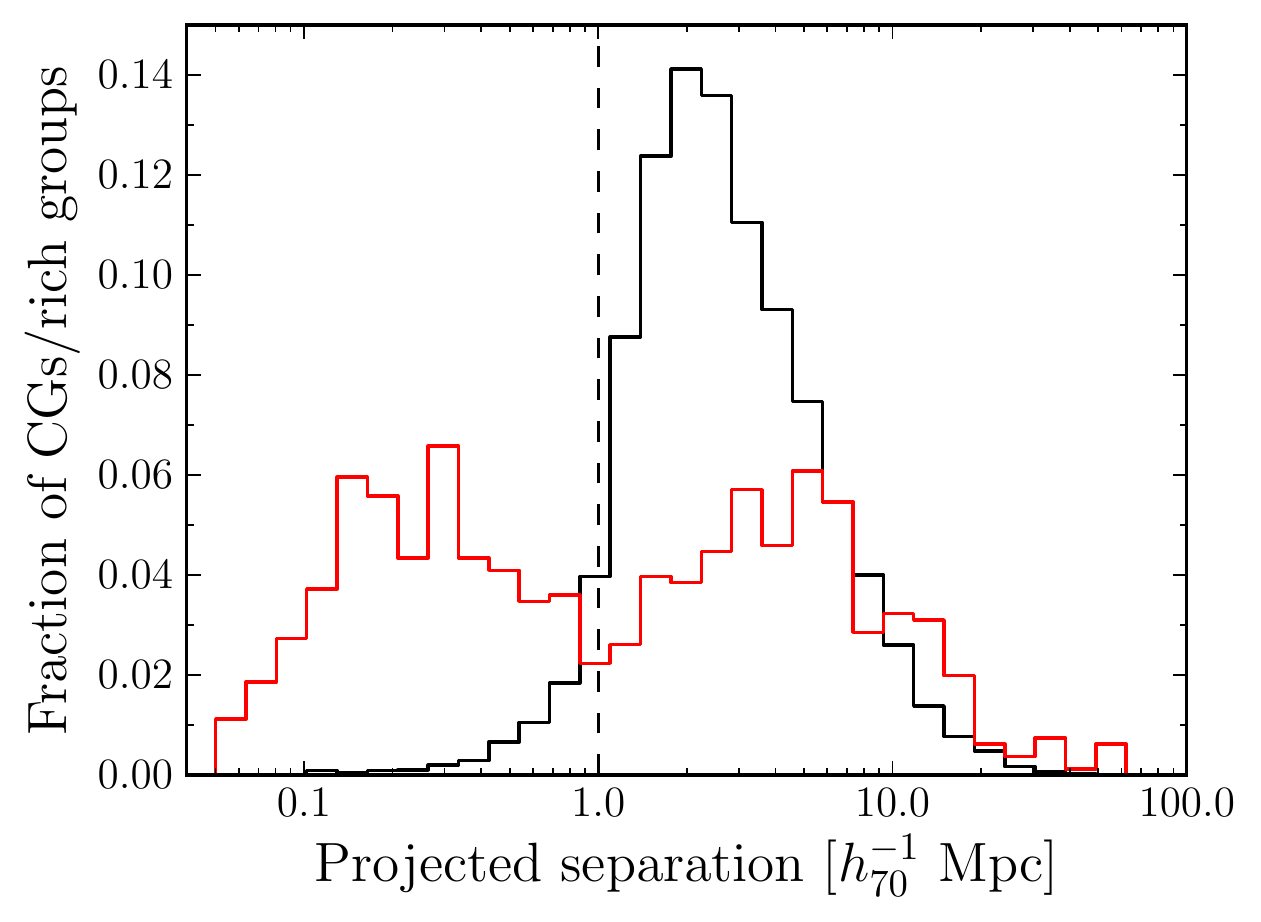}
\caption{Distribution of rich group--CG (red line) and rich group--rich group (black line) projected separations.  Projected separations are given to the nearest rich group within $\Delta z = 0.02$.  The distribution of rich group--CG separations is found to be bimodal, suggesting that CGs can be separated into those residing within the halo of a rich group (left peak) and those more likely associated with structures of comparable richness to the CGs themselves (right peak); the vertical dashed line shows our adopted division between these two CG populations.}
\label{fig:cg_cluster_separation}
\end{figure}

In our observational sample of CGs, we compute the projected separation to the nearest rich group using a fixed redshift interval about the CG centre of $\Delta z = 0.02$ to remove distant foreground or background rich groups.  The results of this calculation are shown Figure \ref{fig:cg_cluster_separation}, where the distribution of rich group--CG separations is shown in red.  Relative to the inter-group separations of rich groups (black histogram in Figure \ref{fig:cg_cluster_separation}), the rich group--CG distribution shows a clear bimodal structure.   Approximately half of CGs are found within 1 Mpc of the nearest rich group, while the remaining CGs are consistent with the distribution of inter-group separations found from the autocorrelation of the T10 catalogue.  The fraction of CGs we find in close proximity to large-scale structure is consistent with that found by \citet{de-carvalho2005} for their sample of CGs, and in excellent agreement with our theoretical results from Paper I.  As discussed in Section \ref{group_incomp}, the T10 catalogue appears to be complete over the richness range we consider here (i.e. $N > 4$), however we cannot rule out that genuinely rich systems are absent from the catalogue of rich groups due to its flux-limited nature.  This effect not withstanding, our results suggest a significant relationship between CGs and rich groups, in that at least half of our CG sample is associated with the richest 10 per cent of structure in the T10 catalogue (see Figure \ref{fig:nmin}). 

The halo model of galaxy clustering \citep[e.g.][and references therein]{cooray2002} generally used to interpret the galaxy autocorrelation function provides a natural framework within which to understand the apparent dichotomy of compact group spatial distributions observed in Figure \ref{fig:cg_cluster_separation}.  In this picture, CGs within $\sim$1 Mpc  of the nearest rich group are embedded within that structure's dark-matter halo; their distribution therefore can be understood as reflecting projected distribution of substructure within these systems.  At wide separations, greater than $\sim$1 Mpc, the distribution of CG--group separations reflects the overall clustering properties of large-scale structure, the so-called halo--halo term in the halo model framework.  Following the discussion in Section \ref{clusters}, our adoption of a richness cut in selecting our rich group comparison sample means the the wide-separation distribution of CGs likely reflects the spatial clustering of poor structures or the CGs themselves.  We use a simple cut of 1 Mpc to differentiate between groups that appear to be hosted within a larger rich group halo -- hereafter referred to as embedded CGs -- and those whose spatial distribution is consistent with that of other distinct physical systems -- hereafter isolated CGs.  This division leads to samples of 413 and 406 isolated and embedded groups containing 1729 and 1726 galaxies, respectively.

\subsection{Galaxy surface number density}
\label{num_dens}

The number-density profiles of groups and clusters are observed to be well described by exponential profiles \citep{carlberg1997,biviano2003}, in agreement with predictions for hierarchical structure formation in a $\Lambda$CDM cosmogony \citep{dubinski1991,navarro1995,navarro1996,navarro1997}.  Studying the number-density profiles of compact groups therefore provides a relatively straight-forward way to assess the physical reality of our distinction between embedded and isolated systems.  Unfortunately, our adopted rich and compact group catalogues are defined in very different ways; In Paper III we use purely photometric criteria while the FoF approach of T10 limits their groups to spectroscopically identified associations.

\begin{figure}
\centering
\includegraphics[scale=0.90]{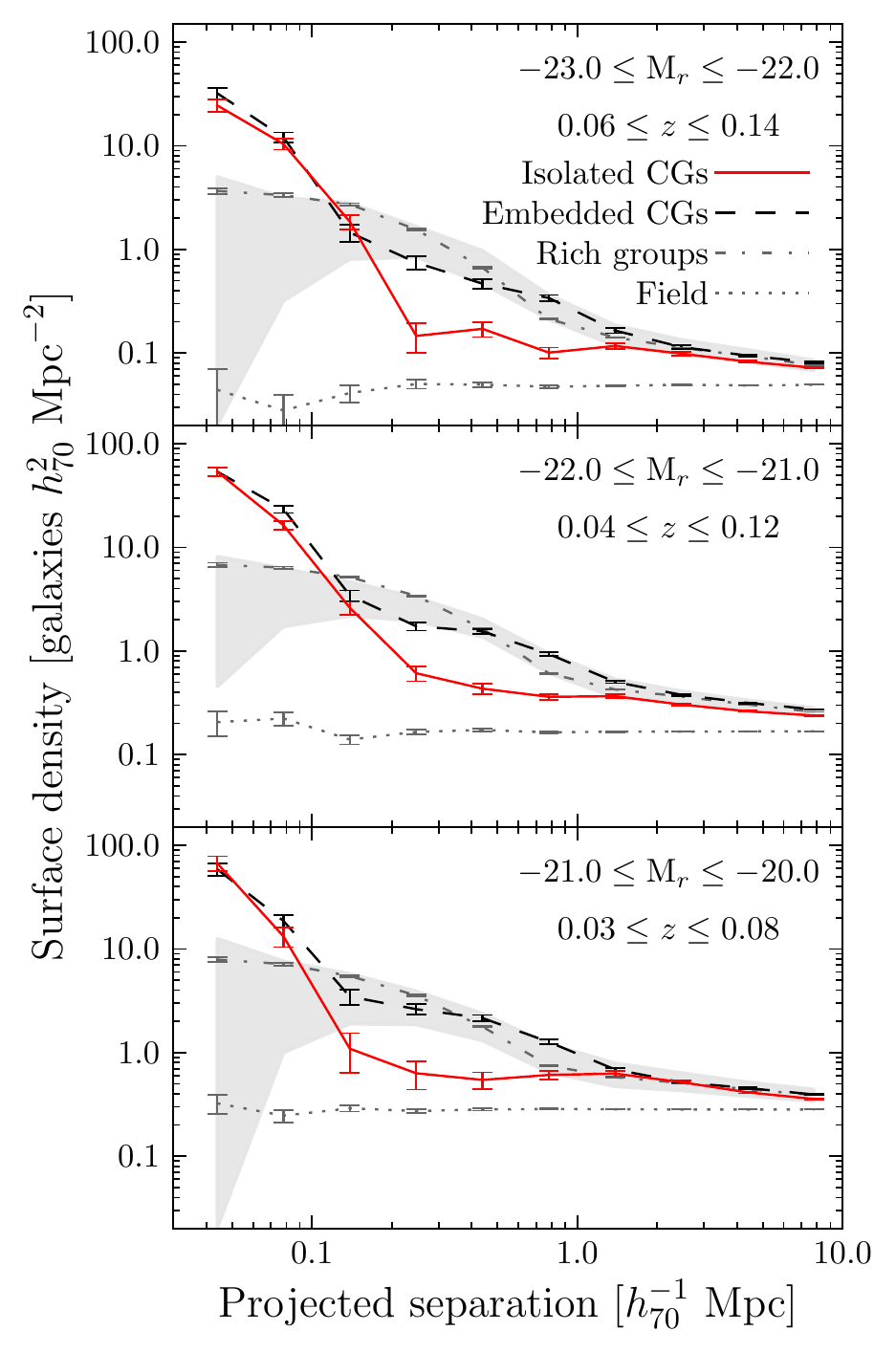}
\caption{Median galaxy surface number density as a function of projected separation from group or CG centre.  Dot-dashed and dotted grey lines represent the number density profile of T10 groups and `field' values measured as described in Section \ref{num_dens}, while solid (red) and dashed (black) lines represent the number-density profiles of isolated and embedded CGs, respectively.  Grey shaded regions represent the range of number densities expected for embedded CGs based solely on their distribution of projected separations from the centres of rich groups (see Section \ref{num_dens} for details). Data are divided such that there are 4 bins per decade in projected separation and Poisson error bars are shown.}
\label{fig:number_density}
\end{figure}

We attempt to standardise our measurements of galaxy number density by adopting only the rich group or CG centres ($\alpha, \delta, z$) and using galaxies in our photometric catalogue to derive galaxy number-density counts in a homogeneous fashion.  In order to account for the redshift-dependent luminosity limit of our SDSS sample we compute several realisations of number density, each using a different volume-limited tracer population.  The results of these measurements are shown in Figure \ref{fig:number_density}, where the three panels correspond to different volume-limited samples with properties as indicated in the upper right of each panel.  We measure galaxy number densities in annuli about the CG or rich group centres, and adopt a fixed line-of-sight binning of $\pm$40 Mpc in order to exclude galaxies outside of the CG or rich group halo.  In addition to the number-density distribution in the volume surrounding CGs and rich groups, we compute a comparison `field' sample using 10\,000 randomly-generated positions within the survey volume, measuring number density in annuli about these field `centres' in the same way as for our CG and rich group samples.  The resulting median field density is shown as a grey dotted line in Figure \ref{fig:number_density}.

The comparison of number-density distributions in Figure \ref{fig:number_density} leads us to two important conclusions.  First, both isolated and embedded CGs (solid and dashed lines, respectively) are characterised by high central number densities, reflecting the explicit selection of high-density systems by the Hickson criteria outlined in Section \ref{cgs}.  Second, the division between isolated and embedded CGs does not appear to be the result of incompleteness in the T10 catalogue; the number-density profile outside the core of isolated CGs -- projected separations $>$200 kpc -- falls off more quickly than either that of rich groups (dot-dashed line in Figure \ref{fig:number_density}) or embedded CGs even when using our volume-limited (i.e. complete) photometric samples.  

The spatial correlation of embedded CGs and rich groups suggests that the number-density profile of embedded CGs should be influenced by their rich group surroundings.  It is therefore interesting to investigate the extent to which embedded CG density profiles can be accounted for {\it entirely} by their association with rich hosts.  We estimate this using Monte Carlo realisations of the embedded CG spatial distribution.  For each realisation, we randomly select $n=406$ rich groups -- where $n$ corresponds to the number of embedded CGs -- and randomly generate positions about these rich groups such that the resulting distribution of projected separations matches the observed distribution of rich group--embedded CG separations (i.e. the distribution of projected separations less than 1 Mpc in Figure \ref{fig:cg_cluster_separation}).  We then compute the number density profiles about these randomly generated positions in annular bins as before, and the shaded regions in Figure \ref{fig:number_density} shows the range of number densities obtained from 1000 realisations of the data.  In all three volume-limited samples, the number-density profile outside of the embedded CG core (projected separations $>$200 kpc) can be accounted for based solely on their relative proximity to rich groups, supporting our interpretation of these CGs as substructure hosted within a larger rich group environment.

\section{Galaxy properties in compact group surroundings}
\label{galaxy_properties}

The distinction between isolated and embedded CGs is rooted in their surrounding environments.  It is therefore interesting to ask if the galaxy populations in and around these systems support our interpretation of embedded and isolated CGs as being fundamentally different.  In this section we discuss the qualitative comparison of two relatively simple characterisations of the galaxy population: the fraction of galaxies on the red sequence and the fraction of disk-dominated (late-type) galaxies.  As before, to provide a consistent comparison across redshift and environment we adopt a volume-limited tracer population for our calculations; however, in this instance we adopt a single volume-limited sample chosen to maximise both the number of galaxies and their magnitude range.  Our adopted sample is defined by $0.056 \le z \le 0.117$ and $-22.7 \le M_r \le -20.9$.

\subsection{Colour distribution}
\label{colour_dist}

There is a strong relationship between galaxy colour and environment such that the fraction of red galaxies in a given population is correlated with local galaxy density \citep{balogh2004,baldry2006}, in good agreement with observations that red galaxies are, in general, more strongly clustered than their blue counterparts \citep[e.g.][]{skibba2009}.  In addition to large-scale correlations, galaxy colour has also proven to be a strong probe of galaxy interactions.  Studies of galaxy pairs have shown that the presence of a nearby companion can often be associated with triggered central star formation, resulting in bluer colours \citep{ellison2010,patton2011}.  Given that the median nearest-neighbour separation in our CG sample is of order the separation where interaction-induced effects are observed to become significant \citep[$\sim$50 kpc;][]{patton2011}, it is interesting to ask if either global or local effects are apparent in the colours of our sample galaxies.

\begin{figure}
\centering
\includegraphics[scale=0.685]{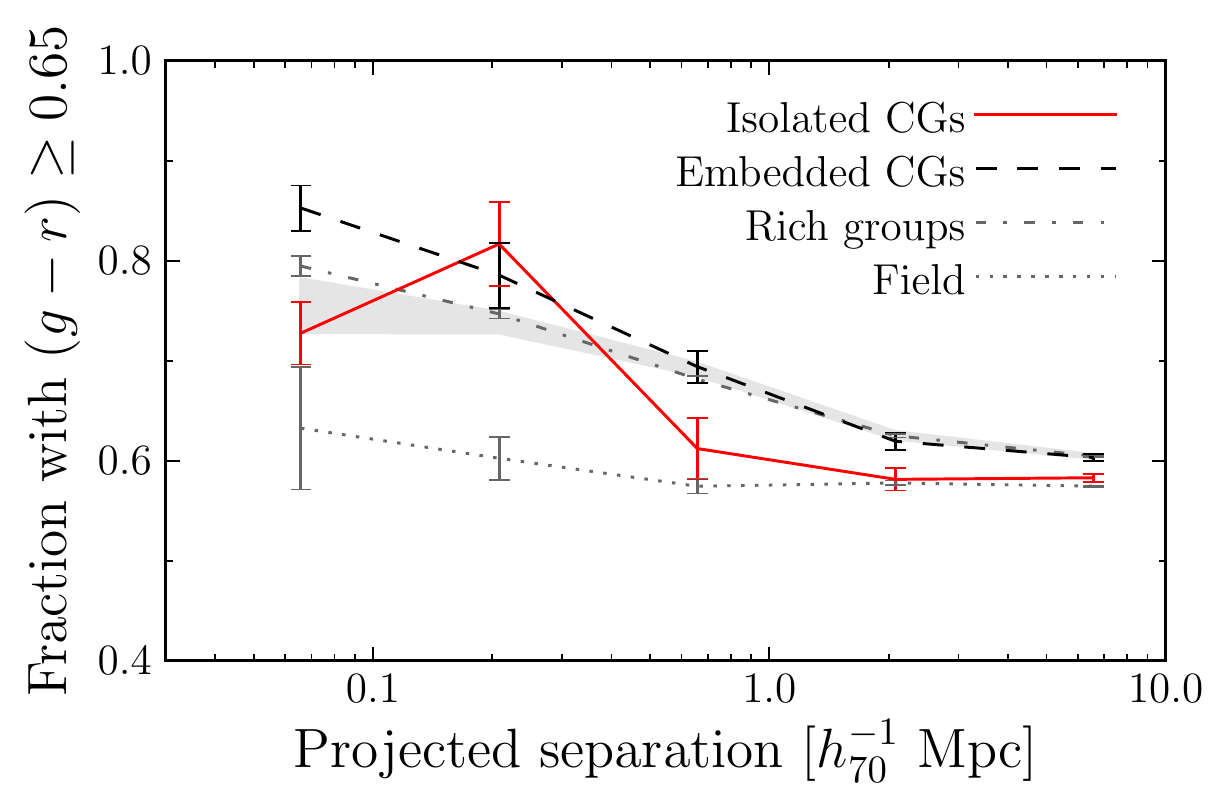}
\caption{Galaxy red fraction as a function of projected separation from group or CG centre.  Red fractions are computed using a colour cut of $(g-r) \ge 0.65$ as described in Section \ref{colour_dist} and include only galaxies with disk axis ratios $b/a > 0.5$ to limit the influence of reddening by edge-on, dusty disks.  Lines and shading are the same as in Figure \ref{fig:number_density}.  Error bars are computed following a binomial distribution and data are divided such that there are 2 bins per decade in projected separation.}
\label{fig:red_fraction}
\end{figure}

We use a fixed rest-frame colour cut of $(g-r) \ge 0.65$~-- chosen based on the bimodal distribution of $g-r$ colours -- to select red galaxies in radial bins of group-centric distance, and include only those galaxies with disk axis ratios $b/a > 0.5$ to limit the influence of edge on disks on galaxy colours.  The resulting relationship between red fraction and radius is shown in Figure \ref{fig:red_fraction}.    There is a clear trend for red fraction to decrease as a function of radius from the group centre in all environments (isolated or embedded CGs and rich groups).  The observed red-sequence fraction in embedded CGs is consistent with that observed for rich groups at all radii and supports our previous conclusion, i.e. that these CGs are embedded within extended, rich systems.  The radial trend of isolated CGs on the other hand, while exhibiting a central red fraction consistent with rich groups (and embedded CGs), shows tentative evidence for a decline outside of the CG core; by 500--600 kpc from the group centre the observed red fraction is consistent with the field value.  

Although noisy, the observed red-fraction trends follow our expectations drawn from previous studies of colour and environment.  Outside of $\sim$200 kpc the red fractions of isolated and embedded CGs diverge in accordance with the galaxy surface densities shown in Figure \ref{fig:number_density}.  Similarly, red fractions measured in the cores of isolated and embedded systems are comparable (within errors), consistent with observations showing that galaxy colour is strongly correlated with local environment, and depends only weakly on the large-scale density field \citep[e.g.][]{blanton2007}.  Further support for this local environment picture can be gained from a combined comparison of galaxy number density and red fraction.  We showed in Figure \ref{fig:number_density} that the median galaxy surface density outside of embedded CG cores can be reproduced based solely on their spatial distribution relative to the centres of rich groups; this is not the case with red fraction, which is systematically higher than expected if it were governed {\it only} by the surrounding, large-scale environment (shown as the grey shaded region in Figure \ref{fig:red_fraction}).

\subsection{Morphological distribution}
\label{morph}

In addition to colour, there is a well known relationship between galaxy morphology and environment such that the fraction of early-type galaxies increases with increasing local galaxy density \citep[e.g.][]{van-der-wel2008}, or alternatively decreasing cluster-centric distance \citep{dressler1980,postman1984}.  As with colour, the comparable central densities and disparate global environments of our isolated and embedded CG samples provide an interesting laboratory to gauge the dependence of morphology on galaxy interactions.

We choose here to characterise morphological distributions based on the fraction of disk-dominated (late-type) galaxies, selected using a cut in $r$-band bulge-to-total luminosity fraction (B/T) of 30 per cent (i.e. B/T$_r < 0.3$).  The reasons for this selection are two-fold.  First, the selection of disk-dominated galaxies based solely on their quantitative B/T is relatively robust \citep{allen2006,cheng2011}, minimising the additional cuts necessary to identify a `clean' morphological tracer from the \citet{simard2011} catalogue.  Second, we expect disks to be heavily affected by their environments.  Both the truncation of star formation in, and tidal disruption of, a galaxy's disk component will lead to rapid evolution toward higher values of B/T, making the fraction of disk light in disk-dominated systems a sensitive probe of such interactions.  As in Figure \ref{fig:red_fraction}, we remove galaxies with disk $b/a < 0.5$ to limit the attenuation of bulge flux by dusty, edge on disks.

\begin{figure}
\centering
\includegraphics[scale=0.685]{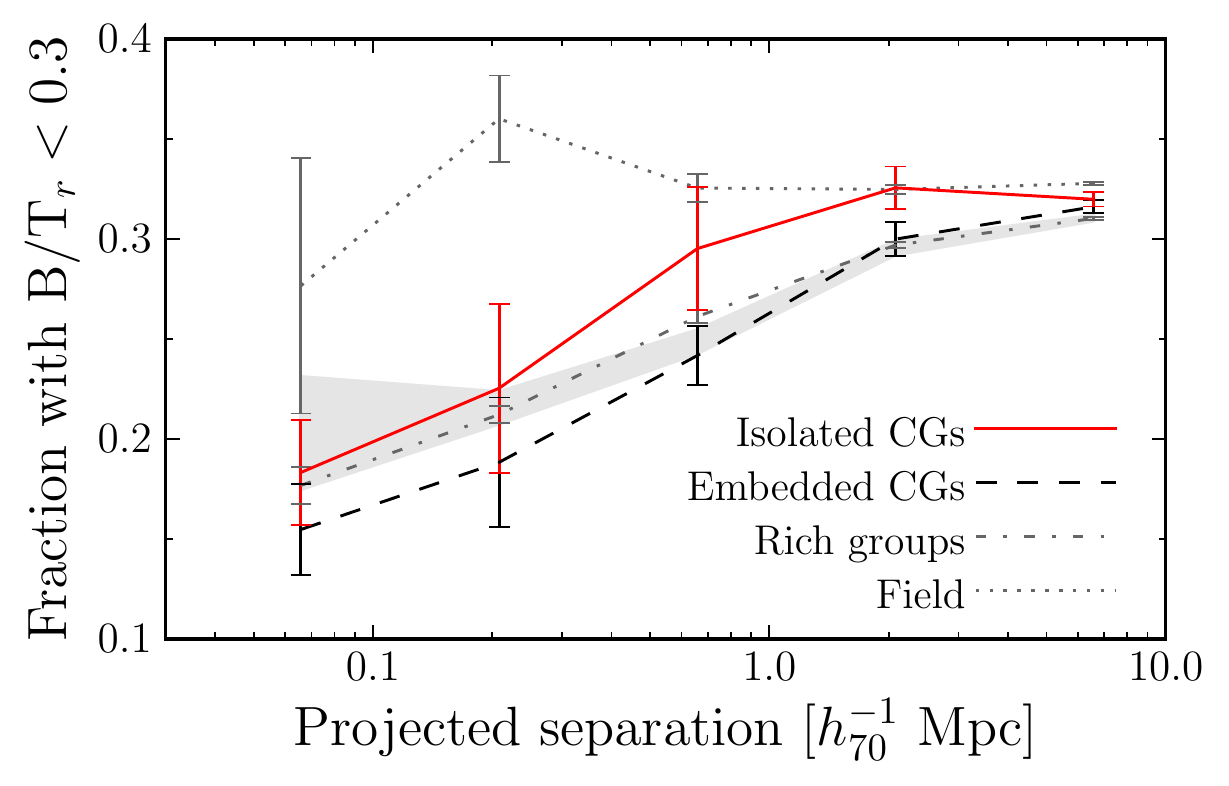}
\caption{Fraction of disk-dominated galaxies as a function of projected separation from group or CG centre.  Only galaxies with disk axis ratios $b/a > 0.5$ are included to limit the attenuation of bulge flux by dusty, edge-on disks.  Lines and shading are the same as previous figures.  Error bars are computed following a binomial distribution and data are divided such that there are 2 bins per decade in projected separation.}
\label{fig:bulge_fraction}
\end{figure}

In Figure \ref{fig:bulge_fraction} we plot the late-type galaxy fraction as a function of radius in each of our environment samples.  In all samples we observe an apparent morphology density relation such that the fraction of late-type, disk-dominated galaxies decreases as a function of decreasing group-centric radius.  The average fraction of disk-dominated galaxies around isolated compact groups appears to increase steeply outside of the CG core, mirroring the red fraction trend shown in Figure \ref{fig:red_fraction}.  In contrast, both embedded CGs and rich groups follow a shallower increase in late-type fraction outside of 200 kpc, commensurate with their comparatively shallow red fraction and number-density radial relations.  Finally, inside $\sim$200 kpc all systems -- that is, isolated and embedded compact groups, as well as rich groups -- host a similar fraction of late-type galaxies despite probing a broad range of both global and local environments.

\begin{figure*}
\centering
\includegraphics[scale=0.9]{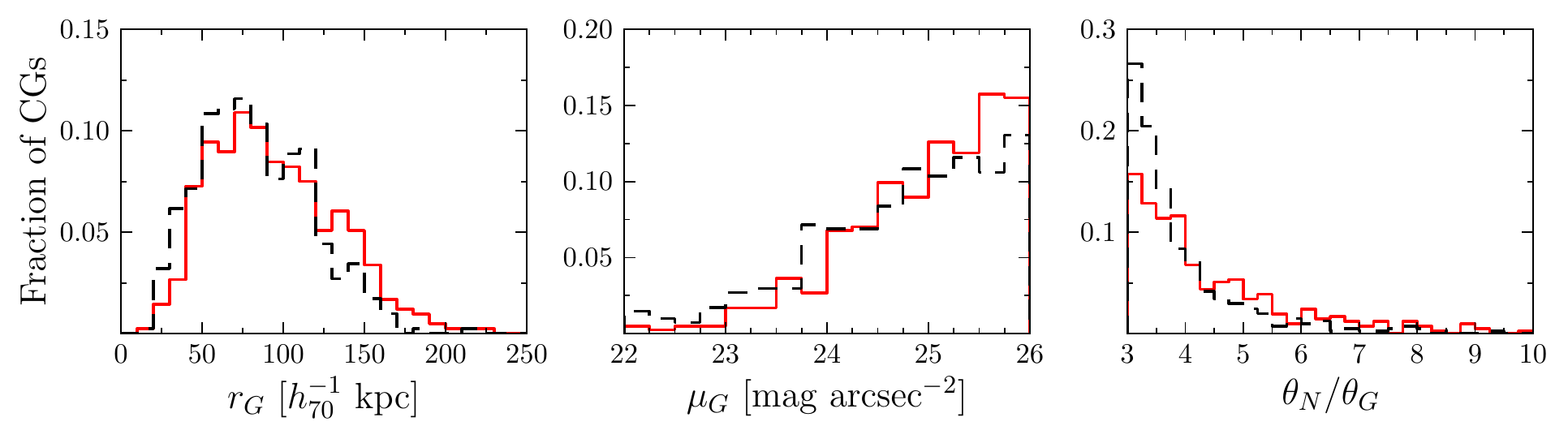}
\caption{Distribution of CG physical size (left panel), surface brightness (middle panel) and relative isolation (right panel).  Distributions for isolated and embedded CGs are shown as solid (red) and dashed (black) histograms, respectively.}
\label{fig:group_prop}
\end{figure*}

\section{Relative properties of isolated and embedded CGs}
\label{cg_properties}

We have so far considered the general properties of galaxies in the volume around CGs without explicitly examining the characteristics of CGs themselves; it is to this comparison that we now turn.  A detailed comparison of CG galaxies with other galaxies in other environments is the focus of a future paper, so here we restrict ourselves to a broad-brush comparison of isolated and embedded CG properties.

\subsection{Group size and surface brightness}

The separation of isolated and embedded CGs based on their relationship to large-scale structure says nothing about the physical properties of the groups themselves; it is therefore interesting to ask if this distinction can be made given the set of parameters that define a compact group, e.g. isolation or surface brightness.

We show in Figure \ref{fig:group_prop} the distributions of group physical size, $r_G$; group surface brightness, $\mu_G$; and relative isolation, $\theta_N/\theta_G$.  Isolated and embedded compact groups are shown as solid (red) and dashed (black) histograms, respectively.  In terms of physical extent, the median size of embedded CGs is $\sim$10 kpc smaller than that of isolated groups (80 versus 90 kpc).  While there is no clear division between the size distributions of isolated and embedded CGs, a two-sample Kolmogorov--Smirnov (KS) shows that the two are unlikely to be drawn from the same parent sample at just under 3$\sigma$ significance ($\sim$0.5 per cent).  The systematic offset between CG sizes is also reflected in their distributions of surface brightness; embedded CGs have on average higher surface brightness, as expected given that $\mu_G \propto \theta_G^{-2}$.  The comparison of surface brightnesses -- and, by extension, sizes -- also suggests that the selecting CGs using cuts in surface brightness \citep[e.g.][]{iovino2003,lee2004} may bias the resulting sample towards predominantly embedded CGs.

From our comparisons with large-scale structure, we know that isolated CGs are distinct from rich groups in terms of their global distribution; the comparison of relative isolation in Figure \ref{fig:group_prop} shows that isolated CGs are also more isolated from other galaxies.  In our sample of SDSS groups, the split between isolated and embedded CGs is roughly equal.  Based on the distribution of relative isolations, we could bias our sample towards isolated CGs using a more strict isolation criteria in our initial group selection.  If we alter criterion (ii) in Section \ref{cgs} to $\theta_N \ge 4\theta_G$, we remove $\sim$50 per cent of embedded CGs (albeit at the expense of 30 per cent of our isolated CG sample).

\subsection{Space density}

Although a direct comparison of CG space density measured from different surveys is difficult, we can nevertheless carry out a qualitative comparison using simple, back-of-the envelope estimates for their frequency following \citet[their equation 1]{lee2004}.  Using our full 1089 CG sample and a median redshift $z=0.1$, we compute a CG space density of 8.9$\times$10$^{-6}$\vmpc; if we restrict ourselves to CGs in the contiguous SDSS Northern Galactic Cap, we find a slightly higher density of 9.9$\times$10$^{-6}$\vmpc.  In their sample of compact groups selected using the \citet{hickson1982} criteria, \citet{mendes-de-oliveira1991} estimate the space density of HCGs to be 1.4$\times$10$^{-5}$\vmpc using simulations; \citet{barton1996} find a similar space density for their spectroscopically-selected CG sample of 1.3$\times$10$^{-5}$\vmpc.  Combined, these disparate samples suggest a relatively redshift-independent space density of HCGs of $\sim$1$\times$10$^{-5}$\vmpc (the median redshifts of the \citeauthor{mendes-de-oliveira1991} and \citeauthor{barton1996} samples are 0.030 and 0.014, respectively).

\citet{lee2004} adopt a more restrictive version of the Hickson selection criteria, requiring $\mu_G < 24$~mag~arcsec$^{-2}$, and find the space density of their SDSS CGs to be 3.9$\times$10$^{-5}$\vmpc, considerably higher than the density estimates for HCGs (using less restrictive criteria).  As discussed in Paper III, \citeauthor{lee2004} use data from the SDSS Early Data Release (EDR),  itself based on a different set of deblending algorithms and photometric calibrations than later data releases; a direct comparison between our CGs and those found by \citeauthor{lee2004} is therefore difficult.  If we consider only the 79 CGs from the EDR that can be identified in the SDSS DR6 (see Paper III), we find a density of 1.8$\times$10$^{-5}$\vmpc, in good agreement with the values quoted above.

In comparison to the \citet{hickson1982} selection criteria, \citet{iovino2003} and \citet{de-carvalho2005} adopt both a more restrictive surface brightness limit -- similar to \citet{lee2004} -- as well as tighter magnitude constraint of $\Delta\mathrm{mag} \le 2$ to limit the influence of projected systems on their group catalogues.  Adopting a median redshift of $z=0.1$ for the 84 CGs of \citeauthor{iovino2003}, we estimate their space density to be 2.8$\times$10$^{-6}$\vmpc, comparable to the estimate obtained for the 459 CG sample of \citet{de-carvalho2005} of 2.9$\times$10$^{-6}$\vmpc (adopting $z=0.12$).  Adopting a surface brightness cut for our sample such that $\mu_G < 24$~mag~arcsec$^{-2}$ results in a space density of $\sim$2$\times$10$^{-6}$\vmpc, in reasonably good agreement (within the uncertainty due to remaining contamination) with results from the \citet{iovino2003} and \citet{de-carvalho2005} samples, suggesting that the factor of 10 lower space density estimated from these samples relative to HCGs are consistent with their more restrictive selection criteria.

\subsection{Galaxy properties}

We have shown that our division of CGs into isolated and embedded is based primarily on environment outside the group core, whereas the comparison of galaxy number-density in Figure \ref{fig:number_density} suggests that inside the core, CG environments are relatively similar.  All other things being equal, we may therefore expect that any differences in the properties of isolated versus embedded CG galaxies are driven by their large-scale environments.

\begin{figure*}
\centering
\includegraphics[scale=0.80]{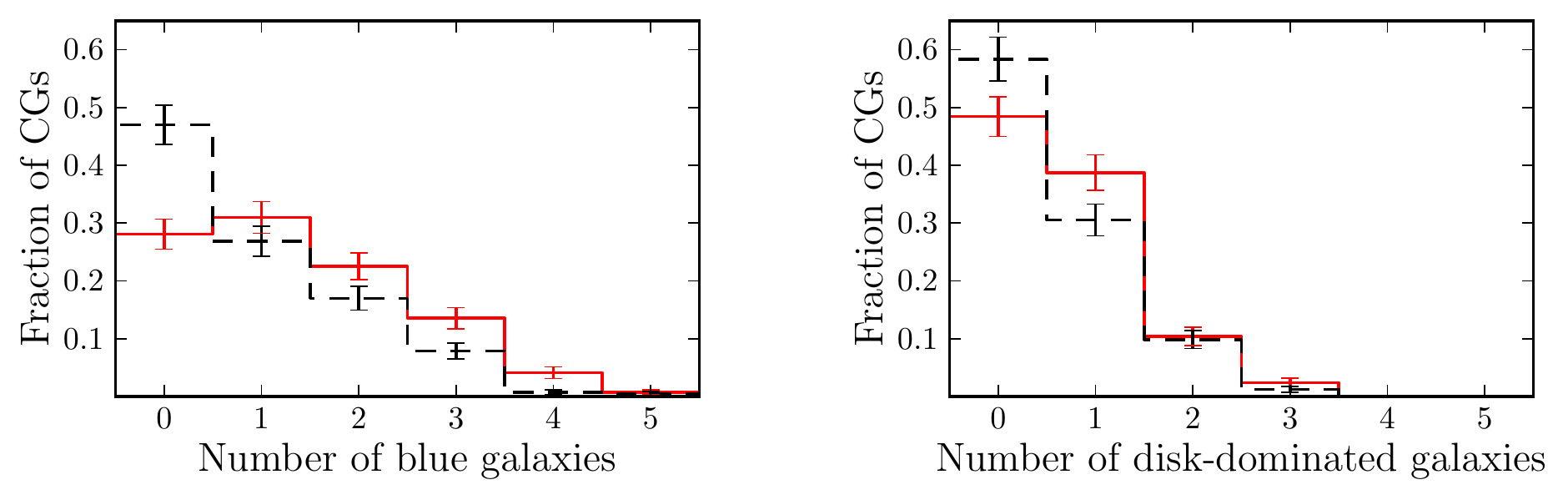}
\caption{Fraction of compact groups hosting a given number of blue galaxies (left) or disk-dominated galaxies (right).  In both panels embedded and isolated CGs are shown by dashed (black) and solid (red) histograms, respectively.  Poisson error bars are shown.}
\label{fig:group_fraction}
\end{figure*}

In Figure \ref{fig:group_fraction} we show the fraction of groups with a given number of blue galaxies (left panel; where blue is defined as $g-r  < 0.65$) or disk-dominated galaxies (right panel) for both embedded and isolated groups (black and red lines, respectively).  We see a clear distinction between isolated and embedded CGs in the colours of their constituent galaxies; $\sim$48 per cent of embedded groups contain no galaxies that we would consider as blue, while this is true for only $\sim$30 per cent of isolated groups.  Isolated CGs, in general, contain a higher number of blue galaxies per group such that the aggregate isolated CG sample has a blue fraction $\sim$10 per cent higher than the embedded CG sample.  Neither sample contains a significant fraction of groups that we might consider to be blue dominated, e.g. three or more blue members; such `blue' systems constitute $\sim$20 per cent of isolated CGs, and only $\sim$10 per cent of embedded CGs.

Turning to morphology, few embedded or isolated CGs ($\sim$10 per cent) contain more than one disk-dominated galaxy.  It is important to note that we have excluded galaxies with disk inclinations less than 60$^{\circ}$ ($b/a < 0.5$) to avoid the effects of edge-on, dusty disks, and therefore the comparison in Figure \ref{fig:group_fraction} represents a lower limit on the true number of disk-dominated galaxies per group.  This caveat notwithstanding, the fraction of groups with 2 or more disk-dominated galaxies is consistent between isolated and embedded systems and suggests that, at the very least, any morphological distinction between these systems comes in galaxies with higher bulge fractions.  

\subsubsection{Effects of residual contamination}
\label{contamination}

In analysing the properties of individual CG galaxies, e.g. Figure \ref{fig:group_fraction}, we must bear in mind that some portion of our cleaned CG sample is made up of projected systems.  In Section \ref{interlopers}, we estimate that our interloper removal scheme reduces the contamination from chance projections by 30 to 35 per cent; however, we must account for the fact that as much as 40 per cent contamination may remain.  It is important to note that the majority of our results up to this point are relatively robust to remaining contamination due to our use of independently-constructed, volume-limited samples.  In this framework, contamination enters our results through the inclusion of spurious group positions but not individual galaxies, and therefore cannot induce the radial trends observed in Figures \ref{fig:number_density}, \ref{fig:red_fraction} and \ref{fig:bulge_fraction} (although it can weaken them).  Unfortunately, the same cannot be said for individual CG galaxies and their properties, the distributions of which are {\it directly} affected by inclusion of contaminating foreground or background galaxies.

\begin{figure*}
\centering
\includegraphics[scale=0.80]{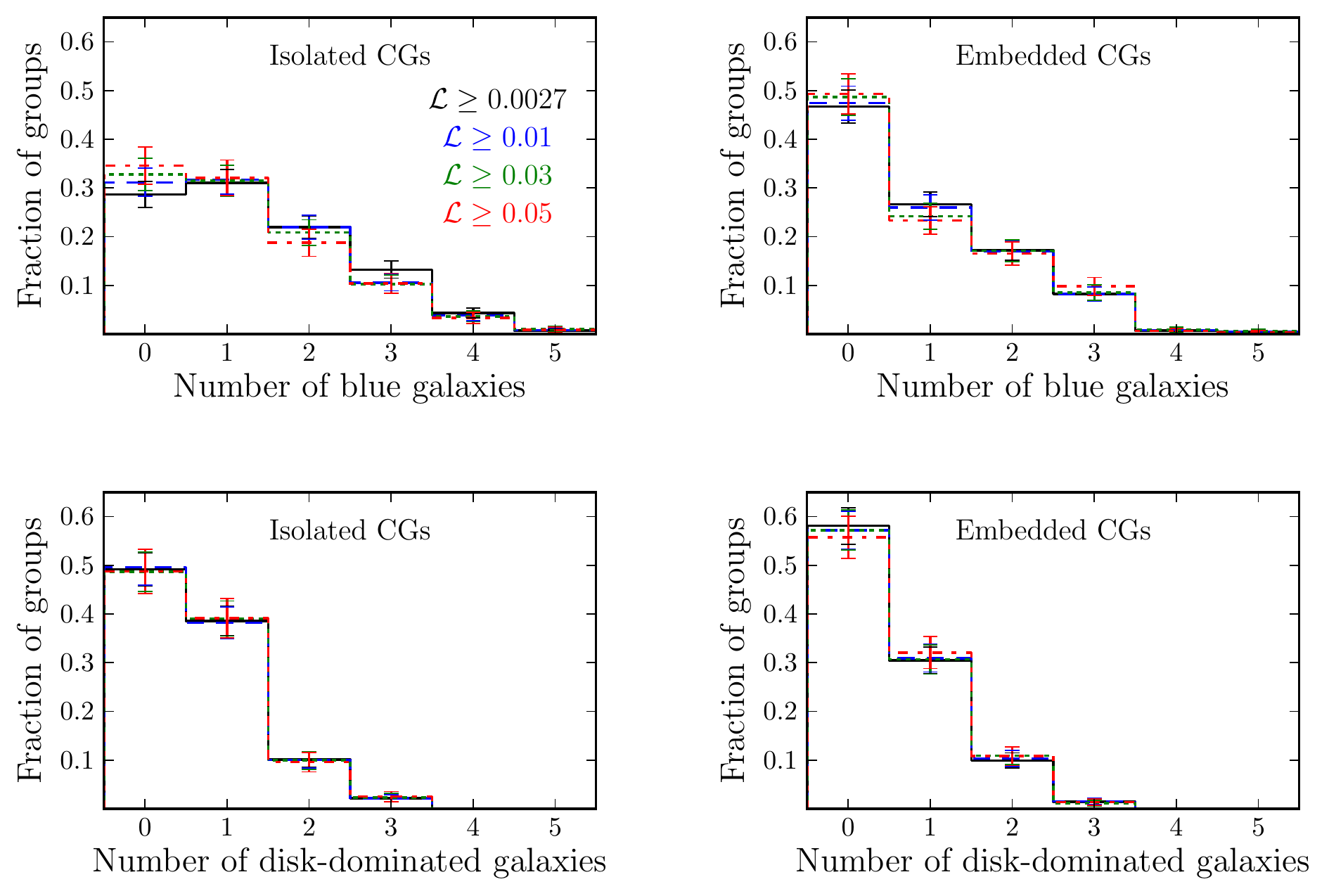}
\caption{The effect of varying likelihood cuts on compact group properties.  Upper panels show the number of blue  galaxies ($g-r < 0.65$), while bottom panels show the number of disk-dominated galaxies (B/T$_r < 0.3$) per group.  Results for different cuts of pairwise likelihood are shown by solid black ($\mathcal{L} \ge$~0.0027), dashed blue ($\mathcal{L} \ge$~0.01), dotted green ($\mathcal{L} \ge$~0.03) and dot-dashed red ($\mathcal{L} \ge$~0.05) histograms. Poisson error bars are shown.}
\label{fig:cont_fraction}
\end{figure*}

We assess the influence of remaining contamination using samples constructed using a variety of pairwise redshift likelihood cuts (see Section \ref{interlopers}) with increasingly stringent tolerance.  In Figure \ref{fig:cont_fraction} we show how the number of blue and disk-dominated galaxies varies for increasingly ``clean'' samples.  Based on these comparisons, we anticipate that residual contamination has only a small effect on our comparison of galaxy properties; even for our most stringent likelihood limit of $\mathcal{L} \ge 0.05$ the fraction of groups in any given bin changes by at most 5 per cent despite an estimated reduction in contamination of 10 to 15 per cent over our baseline cleaned sample (c.f. Figure \ref{fig:contamination}).  We therefore judge that the remaining contamination in our sample has a minimal effect on the inferred properties of CG galaxies.

\section{Summary}

Compact groups offer a unique environment in which to study the influence of dynamical interactions on the galaxy population.  In this paper we use a large, homogeneously-selected sample of CGs from the SDSS DR7 to explore the properties of compact galaxy associations in a large-volume survey.

We have demonstrated that compact groups selected following the criteria of \citet{hickson1982} are divided in their association with large-scale structure; roughly half of CGs can be associated with relatively rich structure, while the remaining half are likely to be either independent structures in the field or associated with comparably poor groups.  The physical reality of this distinction is confirmed by an examination of galaxy number-density profiles, where embedded CGs are found in environments with a factor of 3 to 4 higher galaxy surface density outside $\sim$200 kpc than their isolated counterparts (Figure \ref{fig:number_density}).  Further support for the distinction between CG host environments can be garnered from their surrounding galaxy populations; the observed decline in red-sequence fraction (Figure \ref{fig:red_fraction}) with increasing radius -- or alternatively the increase in late-type galaxy fraction (Figure \ref{fig:bulge_fraction}) -- supports this view, and suggests that isolated CGs are found primarily in field-like environments.

Despite their disparate host environments, the distinction between the galaxy populations of isolated and embedded CGs appears to be relatively small.  We have shown in Section \ref{cg_properties} that isolated groups host a 10 per cent larger blue galaxy population relative to embedded groups, and Figure \ref{fig:group_fraction} shows that this is driven by a global increase in the fraction of blue galaxies hosted per group (rather than a few groups with a 100 per cent blue fraction).  However, the overall fraction of blue galaxies is still relatively low -- $\sim$40 per cent -- even among isolated CGs, suggesting that the high central densities in common between isolated and embedded CGs play a significant role in driving the evolution of their constituent galaxies.  A similar view is gained from consideration of galaxy morphology where, despite hosting a 10 per cent larger disk-dominated galaxy population relative to embedded CGs, the late-type fraction of isolated CGs is still less than $\sim$30 per cent, consistent with previous studies \citep[e.g.][and references therein]{hickson1997}.

The division between isolated and embedded CGs provides a framework within which to understand the formation and evolution of compact groups.  In particular, given the clear association of a significant fraction of the CG population with rich hosts, CGs may hold additional information regarding the processing of galaxies in massive haloes.  It remains the goal of future papers in this series to fully characterise the CG environment and its galaxy population.

\section*{Acknowledgements}
We thank the referee, Roger Coziol, for comments that helped to improve the manuscript.  SLE and DRP gratefully acknowledge the receipt of NSERC Discovery Grants which funded this research.

Funding for the SDSS and SDSS-II has been provided by the Alfred P. Sloan Foundation, the Participating Institutions, the National Science Foundation, the U.S. Department of Energy, the National Aeronautics and Space Administration, the Japanese Monbukagakusho, the Max Planck Society, and the Higher Education Funding Council for England. The SDSS Web Site is http://www.sdss.org/.

The SDSS is managed by the Astrophysical Research Consortium for the Participating Institutions. The Participating Institutions are the American Museum of Natural History, Astrophysical Institute Potsdam, University of Basel, University of Cambridge, Case Western Reserve University, University of Chicago, Drexel University, Fermilab, the Institute for Advanced Study, the Japan Participation Group, Johns Hopkins University, the Joint Institute for Nuclear Astrophysics, the Kavli Institute for Particle Astrophysics and Cosmology, the Korean Scientist Group, the Chinese Academy of Sciences (LAMOST), Los Alamos National Laboratory, the Max-Planck-Institute for Astronomy (MPIA), the Max-Planck-Institute for Astrophysics (MPA), New Mexico State University, Ohio State University, University of Pittsburgh, University of Portsmouth, Princeton University, the United States Naval Observatory, and the University of Washington. 

\bibliographystyle{apj}
\bibliography{biblio}

\begin{thebibliography}{73}
\expandafter\ifx\csname natexlab\endcsname\relax\def\natexlab#1{#1}\fi

\bibitem[{{Abazajian} {et~al.}(2009){Abazajian}, {Adelman-McCarthy},
  {Ag{\"u}eros}, {Allam}, {Allende Prieto}, {An}, {Anderson}, {Anderson},
  {Annis}, {Bahcall}, {Bailer-Jones}, {Barentine}, {Bassett}, {Becker},
  {Beers}, {Bell}, {Belokurov}, {Berlind}, {Berman}, {Bernardi}, {Bickerton},
  {Bizyaev}, {Blakeslee}, {Blanton}, {Bochanski}, {Boroski}, {Brewington},
  {Brinchmann}, {Brinkmann}, {Brunner}, {Budav{\'a}ri}, {Carey}, {Carliles},
  {Carr}, {Castander}, {Cinabro}, {Connolly}, {Csabai}, {Cunha}, {Czarapata},
  {Davenport}, {de Haas}, {Dilday}, {Doi}, {Eisenstein}, {Evans}, {Evans},
  {Fan}, {Friedman}, {Frieman}, {Fukugita}, {G{\"a}nsicke}, {Gates},
  {Gillespie}, {Gilmore}, {Gonzalez}, {Gonzalez}, {Grebel}, {Gunn},
  {Gy{\"o}ry}, {Hall}, {Harding}, {Harris}, {Harvanek}, {Hawley}, {Hayes},
  {Heckman}, {Hendry}, {Hennessy}, {Hindsley}, {Hoblitt}, {Hogan}, {Hogg},
  {Holtzman}, {Hyde}, {Ichikawa}, {Ichikawa}, {Im}, {Ivezi{\'c}}, {Jester},
  {Jiang}, {Johnson}, {Jorgensen}, {Juri{\'c}}, {Kent}, {Kessler}, {Kleinman},
  {Knapp}, {Konishi}, {Kron}, {Krzesinski}, {Kuropatkin}, {Lampeitl},
  {Lebedeva}, {Lee}, {Lee}, {Leger}, {L{\'e}pine}, {Li}, {Lima}, {Lin}, {Long},
  {Loomis}, {Loveday}, {Lupton}, {Magnier}, {Malanushenko}, {Malanushenko},
  {Mandelbaum}, {Margon}, {Marriner}, {Mart{\'{\i}}nez-Delgado}, {Matsubara},
  {McGehee}, {McKay}, {Meiksin}, {Morrison}, {Mullally}, {Munn}, {Murphy},
  {Nash}, {Nebot}, {Neilsen}, {Newberg}, {Newman}, {Nichol}, {Nicinski},
  {Nieto-Santisteban}, {Nitta}, {Okamura}, {Oravetz}, {Ostriker}, {Owen},
  {Padmanabhan}, {Pan}, {Park}, {Pauls}, {Peoples}, {Percival}, {Pier}, {Pope},
  {Pourbaix}, {Price}, {Purger}, {Quinn}, {Raddick}, {Fiorentin}, {Richards},
  {Richmond}, {Riess}, {Rix}, {Rockosi}, {Sako}, {Schlegel}, {Schneider},
  {Scholz}, {Schreiber}, {Schwope}, {Seljak}, {Sesar}, {Sheldon}, {Shimasaku},
  {Sibley}, {Simmons}, {Sivarani}, {Smith}, {Smith}, {Smol{\v c}i{\'c}},
  {Snedden}, {Stebbins}, {Steinmetz}, {Stoughton}, {Strauss}, {Subba Rao},
  {Suto}, {Szalay}, {Szapudi}, {Szkody}, {Tanaka}, {Tegmark}, {Teodoro},
  {Thakar}, {Tremonti}, {Tucker}, {Uomoto}, {Vanden Berk}, {Vandenberg},
  {Vidrih}, {Vogeley}, {Voges}, {Vogt}, {Wadadekar}, {Watters}, {Weinberg},
  {West}, {White}, {Wilhite}, {Wonders}, {Yanny}, {Yocum}, {York}, {Zehavi},
  {Zibetti}, \& {Zucker}}]{abazajian2009}
{Abazajian}, K.~N., {et~al.} 2009, \apjs, 182, 543

\bibitem[{{Aceves} \& {Vel{\'a}zquez}(2002)}]{aceves2002}
{Aceves}, H., \& {Vel{\'a}zquez}, H. 2002, \rmxaa, 38, 199

\bibitem[{{Adelman-McCarthy} {et~al.}(2008){Adelman-McCarthy}, {Ag{\"u}eros},
  {Allam}, {Allende Prieto}, {Anderson}, {Anderson}, {Annis}, {Bahcall},
  {Bailer-Jones}, {Baldry}, {Barentine}, {Bassett}, {Becker}, {Beers}, {Bell},
  {Berlind}, {Bernardi}, {Blanton}, {Bochanski}, {Boroski}, {Brinchmann},
  {Brinkmann}, {Brunner}, {Budav{\'a}ri}, {Carliles}, {Carr}, {Castander},
  {Cinabro}, {Cool}, {Covey}, {Csabai}, {Cunha}, {Davenport}, {Dilday}, {Doi},
  {Eisenstein}, {Evans}, {Fan}, {Finkbeiner}, {Friedman}, {Frieman},
  {Fukugita}, {G{\"a}nsicke}, {Gates}, {Gillespie}, {Glazebrook}, {Gray},
  {Grebel}, {Gunn}, {Gurbani}, {Hall}, {Harding}, {Harvanek}, {Hawley},
  {Hayes}, {Heckman}, {Hendry}, {Hindsley}, {Hirata}, {Hogan}, {Hogg}, {Hyde},
  {Ichikawa}, {Ivezi{\'c}}, {Jester}, {Johnson}, {Jorgensen}, {Juri{\'c}},
  {Kent}, {Kessler}, {Kleinman}, {Knapp}, {Kron}, {Krzesinski}, {Kuropatkin},
  {Lamb}, {Lampeitl}, {Lebedeva}, {Lee}, {Leger}, {L{\'e}pine}, {Lima}, {Lin},
  {Long}, {Loomis}, {Loveday}, {Lupton}, {Malanushenko}, {Malanushenko},
  {Mandelbaum}, {Margon}, {Marriner}, {Mart{\'{\i}}nez-Delgado}, {Matsubara},
  {McGehee}, {McKay}, {Meiksin}, {Morrison}, {Munn}, {Nakajima}, {Neilsen},
  {Newberg}, {Nichol}, {Nicinski}, {Nieto-Santisteban}, {Nitta}, {Okamura},
  {Owen}, {Oyaizu}, {Padmanabhan}, {Pan}, {Park}, {Peoples}, {Pier}, {Pope},
  {Purger}, {Raddick}, {Re Fiorentin}, {Richards}, {Richmond}, {Riess}, {Rix},
  {Rockosi}, {Sako}, {Schlegel}, {Schneider}, {Schreiber}, {Schwope}, {Seljak},
  {Sesar}, {Sheldon}, {Shimasaku}, {Sivarani}, {Smith}, {Snedden}, {Steinmetz},
  {Strauss}, {SubbaRao}, {Suto}, {Szalay}, {Szapudi}, {Szkody}, {Tegmark},
  {Thakar}, {Tremonti}, {Tucker}, {Uomoto}, {Vanden Berk}, {Vandenberg},
  {Vidrih}, {Vogeley}, {Voges}, {Vogt}, {Wadadekar}, {Weinberg}, {West},
  {White}, {Wilhite}, {Yanny}, {Yocum}, {York}, {Zehavi}, \&
  {Zucker}}]{adelman-mccarthy2008}
{Adelman-McCarthy}, J.~K., {et~al.} 2008, \apjs, 175, 297

\bibitem[{{Allen} {et~al.}(2006){Allen}, {Driver}, {Graham}, {Cameron},
  {Liske}, \& {de Propris}}]{allen2006}
{Allen}, P.~D., {Driver}, S.~P., {Graham}, A.~W., {Cameron}, E., {Liske}, J.,
  \& {de Propris}, R. 2006, \mnras, 371, 2

\bibitem[{{Andernach} \& {Coziol}(2005)}]{andernach2005}
{Andernach}, H., \& {Coziol}, R. 2005, in Astronomical Society of the Pacific
  Conference Series, Vol. 329, Nearby Large-Scale Structures and the Zone of
  Avoidance, ed. {A.~P.~Fairall \& P.~A.~Woudt}, 67--76

\bibitem[{{Andernach} \& {Coziol}(2007)}]{andernach2007}
{Andernach}, H., \& {Coziol}, R. 2007, in Groups of Galaxies in the Nearby
  Universe, ed. {I.~Saviane, V.~D.~Ivanov, \& J.~Borissova}, 379--+

\bibitem[{{Athanassoula} {et~al.}(1997){Athanassoula}, {Makino}, \&
  {Bosma}}]{athanassoula1997}
{Athanassoula}, E., {Makino}, J., \& {Bosma}, A. 1997, \mnras, 286, 825

\bibitem[{{Baldry} {et~al.}(2006){Baldry}, {Balogh}, {Bower}, {Glazebrook},
  {Nichol}, {Bamford}, \& {Budavari}}]{baldry2006}
{Baldry}, I.~K., {Balogh}, M.~L., {Bower}, R.~G., {Glazebrook}, K., {Nichol},
  R.~C., {Bamford}, S.~P., \& {Budavari}, T. 2006, \mnras, 373, 469

\bibitem[{{Balogh} {et~al.}(2004){Balogh}, {Baldry}, {Nichol}, {Miller},
  {Bower}, \& {Glazebrook}}]{balogh2004}
{Balogh}, M.~L., {Baldry}, I.~K., {Nichol}, R., {Miller}, C., {Bower}, R., \&
  {Glazebrook}, K. 2004, \apjl, 615, L101

\bibitem[{{Bamford} {et~al.}(2009){Bamford}, {Nichol}, {Baldry}, {Land},
  {Lintott}, {Schawinski}, {Slosar}, {Szalay}, {Thomas}, {Torki}, {Andreescu},
  {Edmondson}, {Miller}, {Murray}, {Raddick}, \& {Vandenberg}}]{bamford2009}
{Bamford}, S.~P., {et~al.} 2009, \mnras, 393, 1324

\bibitem[{{Barnes}(1985)}]{barnes1985}
{Barnes}, J. 1985, \mnras, 215, 517

\bibitem[{{Barton} {et~al.}(1996){Barton}, {Geller}, {Ramella}, {Marzke}, \&
  {da Costa}}]{barton1996}
{Barton}, E., {Geller}, M., {Ramella}, M., {Marzke}, R.~O., \& {da Costa},
  L.~N. 1996, \aj, 112, 871

\bibitem[{{Barton} {et~al.}(1998){Barton}, {de Carvalho}, \&
  {Geller}}]{barton1998}
{Barton}, E.~J., {de Carvalho}, R.~R., \& {Geller}, M.~J. 1998, \aj, 116, 1573

\bibitem[{{Biviano} \& {Girardi}(2003)}]{biviano2003}
{Biviano}, A., \& {Girardi}, M. 2003, \apj, 585, 205

\bibitem[{{Blanton} \& {Berlind}(2007)}]{blanton2007}
{Blanton}, M.~R., \& {Berlind}, A.~A. 2007, \apj, 664, 791

\bibitem[{{Blanton} {et~al.}(2005){Blanton}, {Eisenstein}, {Hogg}, {Schlegel},
  \& {Brinkmann}}]{blanton2005}
{Blanton}, M.~R., {Eisenstein}, D., {Hogg}, D.~W., {Schlegel}, D.~J., \&
  {Brinkmann}, J. 2005, \apj, 629, 143

\bibitem[{{Blanton} \& {Roweis}(2007)}]{blanton2007a}
{Blanton}, M.~R., \& {Roweis}, S. 2007, \aj, 133, 734

\bibitem[{{Brasseur} {et~al.}(2009){Brasseur}, {McConnachie}, {Ellison}, \&
  {Patton}}]{Brasseur2009}
{Brasseur}, C.~M., {McConnachie}, A.~W., {Ellison}, S.~L., \& {Patton}, D.~R.
  2009, \mnras, 392, 1141, {\bf Paper II}

\bibitem[{{Carlberg} {et~al.}(1997){Carlberg}, {Yee}, \&
  {Ellingson}}]{carlberg1997}
{Carlberg}, R.~G., {Yee}, H.~K.~C., \& {Ellingson}, E. 1997, \apj, 478, 462

\bibitem[{{Cheng} {et~al.}(2011){Cheng}, {Faber}, {Simard}, {Graves}, {Lopez},
  {Yan}, \& {Cooper}}]{cheng2011}
{Cheng}, J.~Y., {Faber}, S.~M., {Simard}, L., {Graves}, G.~J., {Lopez}, E.~D.,
  {Yan}, R., \& {Cooper}, M.~C. 2011, \mnras, 412, 727

\bibitem[{{Cooper} {et~al.}(2008){Cooper}, {Tremonti}, {Newman}, \&
  {Zabludoff}}]{cooper2008}
{Cooper}, M.~C., {Tremonti}, C.~A., {Newman}, J.~A., \& {Zabludoff}, A.~I.
  2008, \mnras, 390, 245

\bibitem[{{Cooray} \& {Sheth}(2002)}]{cooray2002}
{Cooray}, A., \& {Sheth}, R. 2002, \physrep, 372, 1

\bibitem[{{Coziol} {et~al.}(2004){Coziol}, {Brinks}, \&
  {Bravo-Alfaro}}]{coziol2004}
{Coziol}, R., {Brinks}, E., \& {Bravo-Alfaro}, H. 2004, \aj, 128, 68

\bibitem[{{Coziol} \& {Plauchu-Frayn}(2007)}]{coziol2007}
{Coziol}, R., \& {Plauchu-Frayn}, I. 2007, \aj, 133, 2630

\bibitem[{{de Carvalho} {et~al.}(2005){de Carvalho}, {Gon{\c c}alves},
  {Iovino}, {Kohl-Moreira}, {Gal}, \& {Djorgovski}}]{de-carvalho2005}
{de Carvalho}, R.~R., {Gon{\c c}alves}, T.~S., {Iovino}, A., {Kohl-Moreira},
  J.~L., {Gal}, R.~R., \& {Djorgovski}, S.~G. 2005, \aj, 130, 425

\bibitem[{{de la Rosa} {et~al.}(2007){de la Rosa}, {de Carvalho}, {Vazdekis},
  \& {Barbuy}}]{de-la-rosa2007}
{de la Rosa}, I.~G., {de Carvalho}, R.~R., {Vazdekis}, A., \& {Barbuy}, B.
  2007, \aj, 133, 330

\bibitem[{{De Lucia} \& {Blaizot}(2007)}]{de-lucia2007}
{De Lucia}, G., \& {Blaizot}, J. 2007, \mnras, 375, 2

\bibitem[{{Diaferio} {et~al.}(1994){Diaferio}, {Geller}, \&
  {Ramella}}]{diaferio1994}
{Diaferio}, A., {Geller}, M.~J., \& {Ramella}, M. 1994, \aj, 107, 868

\bibitem[{{D{\'{\i}}az-Gim{\'e}nez} \& {Mamon}(2010)}]{diaz-gimenez2010}
{D{\'{\i}}az-Gim{\'e}nez}, E., \& {Mamon}, G.~A. 2010, \mnras, 409, 1227

\bibitem[{{Dressler}(1980)}]{dressler1980}
{Dressler}, A. 1980, \apj, 236, 351

\bibitem[{{Dubinski} \& {Carlberg}(1991)}]{dubinski1991}
{Dubinski}, J., \& {Carlberg}, R.~G. 1991, \apj, 378, 496

\bibitem[{{Eke} {et~al.}(2004){Eke}, {Baugh}, {Cole}, {Frenk}, {Norberg},
  {Peacock}, {Baldry}, {Bland-Hawthorn}, {Bridges}, {Cannon}, {Colless},
  {Collins}, {Couch}, {Dalton}, {de Propris}, {Driver}, {Efstathiou}, {Ellis},
  {Glazebrook}, {Jackson}, {Lahav}, {Lewis}, {Lumsden}, {Maddox}, {Madgwick},
  {Peterson}, {Sutherland}, \& {Taylor}}]{eke2004}
{Eke}, V.~R., {et~al.} 2004, \mnras, 348, 866

\bibitem[{{Ellison} {et~al.}(2010){Ellison}, {Patton}, {Simard}, {McConnachie},
  {Baldry}, \& {Mendel}}]{ellison2010}
{Ellison}, S.~L., {Patton}, D.~R., {Simard}, L., {McConnachie}, A.~W.,
  {Baldry}, I.~K., \& {Mendel}, J.~T. 2010, \mnras, 407, 1514

\bibitem[{{Farouki} \& {Shapiro}(1981)}]{farouki1981}
{Farouki}, R., \& {Shapiro}, S.~L. 1981, \apj, 243, 32

\bibitem[{{Goto} {et~al.}(2003){Goto}, {Yamauchi}, {Fujita}, {Okamura},
  {Sekiguchi}, {Smail}, {Bernardi}, \& {Gomez}}]{goto2003}
{Goto}, T., {Yamauchi}, C., {Fujita}, Y., {Okamura}, S., {Sekiguchi}, M.,
  {Smail}, I., {Bernardi}, M., \& {Gomez}, P.~L. 2003, \mnras, 346, 601

\bibitem[{{Governato} {et~al.}(1991){Governato}, {Bhatia}, \&
  {Chincarini}}]{governato1991}
{Governato}, F., {Bhatia}, R., \& {Chincarini}, G. 1991, \apjl, 371, L15

\bibitem[{{Governato} {et~al.}(1996){Governato}, {Tozzi}, \&
  {Cavaliere}}]{governato1996}
{Governato}, F., {Tozzi}, P., \& {Cavaliere}, A. 1996, \apj, 458, 18

\bibitem[{{Gunn} \& {Gott}(1972)}]{gunn1972}
{Gunn}, J.~E., \& {Gott}, III, J.~R. 1972, \apj, 176, 1

\bibitem[{{Hickson}(1982)}]{hickson1982}
{Hickson}, P. 1982, \apj, 255, 382

\bibitem[{{Hickson}(1997)}]{hickson1997}
---. 1997, \araa, 35, 357

\bibitem[{{Hickson} {et~al.}(1992){Hickson}, {Mendes de Oliveira}, {Huchra}, \&
  {Palumbo}}]{hickson1992}
{Hickson}, P., {Mendes de Oliveira}, C., {Huchra}, J.~P., \& {Palumbo}, G.~G.
  1992, \apj, 399, 353

\bibitem[{{Huchra} \& {Geller}(1982)}]{huchra1982}
{Huchra}, J.~P., \& {Geller}, M.~J. 1982, \apj, 257, 423

\bibitem[{{Iovino} {et~al.}(2003){Iovino}, {de Carvalho}, {Gal}, {Odewahn},
  {Lopes}, {Mahabal}, \& {Djorgovski}}]{iovino2003}
{Iovino}, A., {de Carvalho}, R.~R., {Gal}, R.~R., {Odewahn}, S.~C., {Lopes},
  P.~A.~A., {Mahabal}, A., \& {Djorgovski}, S.~G. 2003, \aj, 125, 1660

\bibitem[{{Kauffmann} {et~al.}(2004){Kauffmann}, {White}, {Heckman},
  {M{\'e}nard}, {Brinchmann}, {Charlot}, {Tremonti}, \&
  {Brinkmann}}]{kauffmann2004}
{Kauffmann}, G., {White}, S.~D.~M., {Heckman}, T.~M., {M{\'e}nard}, B.,
  {Brinchmann}, J., {Charlot}, S., {Tremonti}, C., \& {Brinkmann}, J. 2004,
  \mnras, 353, 713

\bibitem[{{Lee} {et~al.}(2004){Lee}, {Allam}, {Tucker}, {Annis}, {Johnston},
  {Scranton}, {Acebo}, {Bahcall}, {Bartelmann}, {B{\"o}hringer}, {Ellman},
  {Grebel}, {Infante}, {Loveday}, {McKay}, {Prada}, {Schneider}, {Stoughton},
  {Szalay}, {Vogeley}, {Voges}, \& {Yanny}}]{lee2004}
{Lee}, B.~C., {et~al.} 2004, \aj, 127, 1811

\bibitem[{{Mart{\'{\i}}nez} {et~al.}(2008){Mart{\'{\i}}nez}, {del Olmo},
  {Coziol}, \& {Focardi}}]{martinez2008}
{Mart{\'{\i}}nez}, M.~A., {del Olmo}, A., {Coziol}, R., \& {Focardi}, P. 2008,
  \apjl, 678, L9

\bibitem[{{McConnachie} {et~al.}(2008){McConnachie}, {Ellison}, \&
  {Patton}}]{mcconnachie2008}
{McConnachie}, A.~W., {Ellison}, S.~L., \& {Patton}, D.~R. 2008, \mnras, 387,
  1281, {\bf Paper I}

\bibitem[{{McConnachie} {et~al.}(2009){McConnachie}, {Patton}, {Ellison}, \&
  {Simard}}]{mcconnachie2009}
{McConnachie}, A.~W., {Patton}, D.~R., {Ellison}, S.~L., \& {Simard}, L. 2009,
  \mnras, 395, 255, {\bf Paper III}

\bibitem[{{McIntosh} {et~al.}(2008){McIntosh}, {Guo}, {Hertzberg}, {Katz},
  {Mo}, {van den Bosch}, \& {Yang}}]{mcintosh2008}
{McIntosh}, D.~H., {Guo}, Y., {Hertzberg}, J., {Katz}, N., {Mo}, H.~J., {van
  den Bosch}, F.~C., \& {Yang}, X. 2008, \mnras, 388, 1537

\bibitem[{{Mendes de Oliveira} \& {Hickson}(1991)}]{mendes-de-oliveira1991}
{Mendes de Oliveira}, C., \& {Hickson}, P. 1991, \apj, 380, 30

\bibitem[{{Mendes de Oliveira} \& {Hickson}(1994)}]{mendes-de-oliveira1994}
---. 1994, \apj, 427, 684

\bibitem[{{Moore} {et~al.}(1996){Moore}, {Katz}, {Lake}, {Dressler}, \&
  {Oemler}}]{moore1996}
{Moore}, B., {Katz}, N., {Lake}, G., {Dressler}, A., \& {Oemler}, A. 1996,
  \nat, 379, 613

\bibitem[{{Navarro} {et~al.}(1995){Navarro}, {Frenk}, \& {White}}]{navarro1995}
{Navarro}, J.~F., {Frenk}, C.~S., \& {White}, S.~D.~M. 1995, \mnras, 275, 56

\bibitem[{{Navarro} {et~al.}(1996){Navarro}, {Frenk}, \& {White}}]{navarro1996}
---. 1996, \apj, 462, 563

\bibitem[{{Navarro} {et~al.}(1997){Navarro}, {Frenk}, \& {White}}]{navarro1997}
---. 1997, \apj, 490, 493

\bibitem[{{Niemi} {et~al.}(2007){Niemi}, {Nurmi}, {Hein{\"a}m{\"a}ki}, \&
  {Valtonen}}]{niemi2007}
{Niemi}, S., {Nurmi}, P., {Hein{\"a}m{\"a}ki}, P., \& {Valtonen}, M. 2007,
  \mnras, 382, 1864

\bibitem[{{Nulsen}(1982)}]{nulsen1982}
{Nulsen}, P.~E.~J. 1982, \mnras, 198, 1007

\bibitem[{{Palumbo} {et~al.}(1995){Palumbo}, {Saracco}, {Hickson}, \& {Mendes
  de Oliveira}}]{palumbo1995}
{Palumbo}, G.~G.~C., {Saracco}, P., {Hickson}, P., \& {Mendes de Oliveira}, C.
  1995, \aj, 109, 1476

\bibitem[{{Patton} {et~al.}(2011){Patton}, {Ellison}, {Simard}, {McConnachie},
  \& {Mendel}}]{patton2011}
{Patton}, D.~R., {Ellison}, S.~L., {Simard}, L., {McConnachie}, A.~W., \&
  {Mendel}, J.~T. 2011, \mnras, 412, 591

\bibitem[{{Postman} \& {Geller}(1984)}]{postman1984}
{Postman}, M., \& {Geller}, M.~J. 1984, \apj, 281, 95

\bibitem[{{Proctor} {et~al.}(2004){Proctor}, {Forbes}, {Hau}, {Beasley}, {De
  Silva}, {Contreras}, \& {Terlevich}}]{proctor2004a}
{Proctor}, R.~N., {Forbes}, D.~A., {Hau}, G.~K.~T., {Beasley}, M.~A., {De
  Silva}, G.~M., {Contreras}, R., \& {Terlevich}, A.~I. 2004, \mnras, 349, 1381

\bibitem[{{Ramella} {et~al.}(1994){Ramella}, {Diaferio}, {Geller}, \&
  {Huchra}}]{ramella1994}
{Ramella}, M., {Diaferio}, A., {Geller}, M.~J., \& {Huchra}, J.~P. 1994, \aj,
  107, 1623

\bibitem[{{Ramella} {et~al.}(1997){Ramella}, {Pisani}, \&
  {Geller}}]{ramella1997}
{Ramella}, M., {Pisani}, A., \& {Geller}, M.~J. 1997, \aj, 113, 483

\bibitem[{{Rood} \& {Struble}(1994)}]{rood1994}
{Rood}, H.~J., \& {Struble}, M.~F. 1994, \pasp, 106, 413

\bibitem[{{Rose}(1977)}]{rose1977}
{Rose}, J.~A. 1977, \apj, 211, 311

\bibitem[{{Simard} {et~al.}(2011){Simard}, {Mendel}, {Patton}, {Ellison}, \&
  {McConnachie}}]{simard2011}
{Simard}, L., {Mendel}, J.~T., {Patton}, D.~R., {Ellison}, S.~L., \&
  {McConnachie}, A.~W. 2011, \apjs, submitted

\bibitem[{{Simard} {et~al.}(2002){Simard}, {Willmer}, {Vogt}, {Sarajedini},
  {Phillips}, {Weiner}, {Koo}, {Im}, {Illingworth}, \& {Faber}}]{simard2002}
{Simard}, L., {et~al.} 2002, \apjs, 142, 1

\bibitem[{{Skibba} {et~al.}(2009){Skibba}, {Bamford}, {Nichol}, {Lintott},
  {Andreescu}, {Edmondson}, {Murray}, {Raddick}, {Schawinski}, {Slosar},
  {Szalay}, {Thomas}, \& {Vandenberg}}]{skibba2009}
{Skibba}, R.~A., {et~al.} 2009, \mnras, 399, 966

\bibitem[{{Springel} {et~al.}(2005){Springel}, {White}, {Jenkins}, {Frenk},
  {Yoshida}, {Gao}, {Navarro}, {Thacker}, {Croton}, {Helly}, {Peacock}, {Cole},
  {Thomas}, {Couchman}, {Evrard}, {Colberg}, \& {Pearce}}]{springel2005}
{Springel}, V., {et~al.} 2005, \nat, 435, 629

\bibitem[{{Tago} {et~al.}(2010){Tago}, {Saar}, {Tempel}, {Einasto}, {Einasto},
  {Nurmi}, \& {Hein{\"a}m{\"a}ki}}]{tago2010}
{Tago}, E., {Saar}, E., {Tempel}, E., {Einasto}, J., {Einasto}, M., {Nurmi},
  P., \& {Hein{\"a}m{\"a}ki}, P. 2010, \aap, 514, A102+, {\bf T10}

\bibitem[{{Toomre} \& {Toomre}(1972)}]{toomre1972}
{Toomre}, A., \& {Toomre}, J. 1972, \apj, 178, 623

\bibitem[{{van der Wel}(2008)}]{van-der-wel2008}
{van der Wel}, A. 2008, \apjl, 675, L13

\bibitem[{{Zepf} {et~al.}(1991){Zepf}, {Whitmore}, \& {Levison}}]{zepf1991a}
{Zepf}, S.~E., {Whitmore}, B.~C., \& {Levison}, H.~F. 1991, \apj, 383, 524

\end{thebibliography}

\appendix
\section{Cleaned CG catalogues}
\label{tables}

Tables \ref{tab:clean_groups} and \ref{tab:clean_galaxies} list the properties of compact groups initially presented in Paper III and subsequently `cleaned' following the interloper rejection procedure described in Section \ref{interlopers}.  Only groups that pass the cleaning procedure are included.  We only show group information for the first 20 CGs (Table \ref{tab:clean_groups}) and galaxy information for the first 5 groups (Table \ref{tab:clean_galaxies}); full tables are available online.

Table \ref{tab:clean_groups} lists the properties for each group, with columns as follows:\\

\indent {\it Column 1} -- Group ID in the CG catalogue of \citet{mcconnachie2009}.\\
\indent {\it Columns 2, 3, 4} -- Right Ascension (J2000) of the geometric group centre. \\
\indent {\it Columns 5, 6, 7} -- Declination (J2000) of the geometric group centre. \\
\indent {\it Column 8} -- Estimated redshift of the group, determined from the joint probability of group member redshifts.\\
\indent {\it Column 9} -- Estimated uncertainty on the group redshift, determined from the 1$\sigma$ width of the joint redshift probability.\\
\indent {\it Column 10} -- Number of group members in the original \citet{mcconnachie2009} photometric group catalogue (i.e. prior to cleaning).\\
\indent {\it Column 11} -- Number of remaining group members after application of the cleaning algorithm described in Section \ref{interlopers}.\\
\indent {\it Column 12} -- Distance to the nearest rich group in the \citet{tago2010} catalogue in \ih~Mpc, estimated following the description in Section \ref{corr}.\\

Table \ref{tab:clean_galaxies} lists the properties for each galaxy, with columns as follows:\\

\indent {\it Column 1} -- Galaxy ID in the group catalogue of \citet{mcconnachie2009}.  Formatting is {\it groupID.galaxy}, where {\it groupID} corresponds to the Group ID in Table \ref{tab:clean_groups} and {\it galaxy} has a value $i = 1,...,n_\mathrm{M09}$, corresponding to the galaxies number within the group. \\
\indent{\it Column 2} -- Galaxy object ID within the SDSS. \\
\indent {\it Columns 3, 4, 5} -- Right Ascension (J2000) of the galaxy. \\
\indent {\it Columns 6, 7, 8} -- Declination (J2000) of the galaxy. \\
\indent {\it Columns 9, 10} -- Spectroscopic redshift and its corresponding uncertainty, when available.\\
\indent {\it Columns 11, 12} -- Photometric redshift and its corresponding uncertainty.  Photometric redshifts from the SDSS {\tt Photoz} table are given.\\

\begin{table*}
\centering
\begin{tabular}{@{}cccccccccccc@{}} \toprule
ID & \multicolumn{3}{c}{$\alpha$~[J2000]} & \multicolumn{3}{c}{$\delta$~[J2000]} & $z_{\mathrm{est.}}$ & $\Delta z_\mathrm{est.}$ & $n_\mathrm{M09}$ & $n_\mathrm{clean}$ & $R_\mathrm{T10}$~[\ih~Mpc] \\
(1) & (2) & (3) & (4) & (5) & (6) & (7) & (8) & (9) & (10) & (11) & (12) \\ \midrule
SDSSCGA00390 & 14 & 28 & 30.48 & +20 & 15 & 52.56 & 0.124 & 0.0001 & 4 & 4 & 2.03 \\
SDSSCGA02235 & 13 & 27 & 50.40 & +35 & 23 & 19.32 & 0.154 & 0.0001 & 5 & 5 &   \\
SDSSCGA02239 & 14 & 16 & 01.44 & +06 & 04 & 48.10 & 0.131 & 0.0001 & 6 & 5 & 11.69 \\
SDSSCGA01831 & 15 & 35 & 57.84 & +51 & 25 & 39.72 & 0.160 & 0.0001 & 4 & 4 &   \\
SDSSCGA01832 & 09 & 19 & 22.80 & +56 & 32 & 37.68 & 0.122 & 0.0001 & 5 & 5 & 7.09 \\
SDSSCGA01833 & 14 & 37 & 02.40 & +08 & 03 & 22.71 & 0.059 & 0.0001 & 4 & 4 & 0.30 \\
SDSSCGA01836 & 10 & 42 & 30.00 & +57 & 17 & 38.04 & 0.073 & 0.0001 & 5 & 4 & 1.30 \\
SDSSCGA01839 & 13 & 04 & 29.52 & +24 & 36 & 38.16 & 0.147 & 0.0001 & 4 & 4 &   \\
SDSSCGA02099 & 10 & 39 & 36.48 & +39 & 28 & 57.00 & 0.117 & 0.0001 & 5 & 5 & 6.65 \\
SDSSCGA02098 & 10 & 05 & 07.20 & +14 & 05 & 09.24 & 0.085 & 0.0001 & 6 & 6 & 0.14 \\
SDSSCGA02093 & 13 & 20 & 43.68 & +57 & 16 & 17.40 & 0.117 & 0.0001 & 5 & 4 & 0.17 \\
SDSSCGA02092 & 08 & 16 & 48.00 & +08 & 09 & 50.79 & 0.052 & 0.0001 & 4 & 4 & 6.11 \\
SDSSCGA02091 & 17 & 03 & 17.76 & +63 & 04 & 04.43 & 0.083 & 0.0001 & 4 & 4 &   \\
SDSSCGA01693 & 10 & 00 & 17.28 & +61 & 57 & 55.80 & 0.134 & 0.0002 & 4 & 4 & 6.10 \\
SDSSCGA02097 & 10 & 24 & 15.84 & +06 & 55 & 18.01 & 0.052 & 0.0001 & 4 & 4 & 1.62 \\
SDSSCGA01947 & 11 & 10 & 24.24 & +34 & 27 & 41.40 & 0.079 & 0.0001 & 5 & 4 & 0.66 \\
SDSSCGA01940 & 10 & 41 & 57.36 & +15 & 45 & 18.00 & 0.051 & 0.0002 & 4 & 4 & 0.73 \\
SDSSCGA01943 & 08 & 38 & 39.12 & +02 & 48 & 44.10 & 0.170 & 0.0055 & 5 & 4 &   \\
SDSSCGA00942 & 13 & 48 & 23.52 & +06 & 33 & 40.39 & 0.082 & 0.0001 & 4 & 4 & 3.29 \\
SDSSCGA00945 & 12 & 02 & 49.20 & +10 & 40 & 11.64 & 0.060 & 0.0001 & 4 & 4 & 2.91 \\
\bottomrule
\end{tabular}
\caption{Compact groups that pass the interloper removal procedure described in Section \ref{interlopers}.  Only the first 20 groups are shown; the full table contains 1086 rows.  See Appendix \ref{tables} for a description of each column.}
\label{tab:clean_groups}
\end{table*}

\begin{table*}
\centering
\begin{tabular}{@{}cccccccccccc@{}} \toprule
ID & SDSS objID & \multicolumn{3}{c}{$\alpha$~[J2000]} & \multicolumn{3}{c}{$\delta$~[J2000]} & $z_{\mathrm{spec}}$ & $\Delta z_\mathrm{spec}$ & $z_{\mathrm{photo}}$ & $\Delta z_{\mathrm{photo}}$ \\ 
(1) & (2) & (3) & (4) & (5) & (6) & (7) & (8) & (9) & (10) & (11) & (12) \\ \midrule
SDSSCGA00390.1 & 587742062696726690 & 14 & 28 & 31.92 & +20 & 16 & 12.36 & 0.123 & 0.0001 & 0.119 & 0.0202 \\
SDSSCGA00390.2 & 587742062696661218 & 14 & 28 & 29.28 & +20 & 16 & 00.84 &  &  & 0.158 & 0.0146 \\
SDSSCGA00390.3 & 587742062696661231 & 14 & 28 & 30.72 & +20 & 15 & 38.52 & 0.126 & 0.0001 & 0.106 & 0.0133 \\
SDSSCGA00390.4 & 587742062696661232 & 14 & 28 & 30.00 & +20 & 15 & 38.88 &  &  & 0.142 & 0.0142 \\
SDSSCGA02235.1 & 587739130803716242 & 13 & 27 & 54.96 & +35 & 23 & 35.88 &  &  & 0.159 & 0.0128 \\
SDSSCGA02235.2 & 587739130803716218 & 13 & 27 & 46.56 & +35 & 23 & 04.20 &  &  & 0.159 & 0.0095 \\
SDSSCGA02235.3 & 587739130803716217 & 13 & 27 & 48.00 & +35 & 22 & 53.40 &  &  & 0.165 & 0.0132 \\
SDSSCGA02235.4 & 587739130803716231 & 13 & 27 & 49.92 & +35 & 22 & 54.84 & 0.154 & 0.0001 & 0.173 & 0.0187 \\
SDSSCGA02235.5 & 587739130803716234 & 13 & 27 & 52.32 & +35 & 24 & 07.92 & 0.153 & 0.0001 & 0.151 & 0.0098 \\
SDSSCGA02239.1 & 587736524299108484 & 14 & 15 & 58.56 & +06 & 04 & 36.91 & 0.132 & 0.0002 & 0.133 & 0.0080 \\
SDSSCGA02239.2 & 587736524299108646 & 14 & 16 & 00.96 & +06 & 05 & 32.92 & 0.132 & 0.0002 & 0.126 & 0.0106 \\
SDSSCGA02239.3 & 587736524299108659 & 14 & 16 & 05.28 & +06 & 05 & 44.95 & 0.130 & 0.0001 & 0.134 & 0.0109 \\
SDSSCGA02239.4 & 587736524299108624 & 14 & 15 & 56.64 & +06 & 04 & 30.72 &  &  & 0.144 & 0.0109 \\
SDSSCGA02239.5 & 587736524299108724 & 14 & 16 & 05.04 & +06 & 04 & 08.54 & 0.131 & 0.0001 & 0.122 & 0.0270 \\
SDSSCGA01831.1 & 588011102104912065 & 15 & 35 & 57.60 & +51 & 25 & 30.36 & 0.158 & 0.0002 & 0.171 & 0.0119 \\
SDSSCGA01831.2 & 588011102104912064 & 15 & 35 & 56.64 & +51 & 25 & 17.04 & 0.162 & 0.0002 & 0.155 & 0.0088 \\
SDSSCGA01831.3 & 588011102104912067 & 15 & 36 & 00.48 & +51 & 25 & 14.16 &  &  & 0.155 & 0.0258 \\
SDSSCGA01831.4 & 588011102104912049 & 15 & 35 & 56.40 & +51 & 26 & 36.60 &  &  & 0.142 & 0.0166 \\
SDSSCGA01832.1 & 587725470136074423 & 09 & 19 & 25.44 & +56 & 32 & 41.64 &  &  & 0.132 & 0.0308 \\
SDSSCGA01832.2 & 587725470136074424 & 09 & 19 & 25.68 & +56 & 32 & 48.48 & 0.122 & 0.0001 & 0.167 & 0.0192 \\
SDSSCGA01832.3 & 587725470136074426 & 09 & 19 & 24.96 & +56 & 32 & 40.92 &  &  & 0.248 & 0.0938 \\
SDSSCGA01832.4 & 587725470136074414 & 09 & 19 & 18.48 & +56 & 32 & 16.08 &  &  & 0.110 & 0.0215 \\
SDSSCGA01832.5 & 587725470136074421 & 09 & 19 & 20.16 & +56 & 32 & 42.72 &  &  & 0.109 & 0.0114 \\
\bottomrule
\end{tabular}
\caption{Individual member galaxies of the groups in Table \ref{tab:clean_groups}.  Only information for the first 5 groups is shown; the full table contains 4566 rows.  See Appendix \ref{tables} for a description of each column.}
\label{tab:clean_galaxies}
\end{table*}


\end{document}